\documentclass[apj]{emulateapj}
\usepackage{lscape}

\shorttitle{Abundances from Medium Resolution Spectroscopy}
\shortauthors{Kirby et al.}
\begin{document}
\newcommand{\deep}{\textsc{deep2}}
\newcommand{\deimos}{\textsc{deimos}}
\newcommand{\cfht}{\textsc{cfht}}
\newcommand{\ha}{H$\alpha$}
\newcommand{\teff}{$T_{\mathrm{eff}}$}
\newcommand{\mathteff}{T_{\mathrm{eff}}}
\newcommand{\logg}{$\log g$}
\newcommand{\mathlogg}{\log g}
\newcommand{\feh}{[Fe/H]}
\newcommand{\mathfeh}{\mathrm{[Fe/H]}}
\newcommand{\afe}{[$\alpha$/Fe]}
\newcommand{\mathafe}{\mathrm{[\alpha/Fe]}}
\newcommand{\vt}{$v_t$}
\newcommand{\mathaa}{\mathrm{\AA}}
\newcommand{\ngcaf}{\object[M79]{NGC~1904 (M79)}}
\newcommand{\ngca}{\object{M79}}
\newcommand{\ngcb}{\object{NGC 2419}}
\newcommand{\ngccf}{\object[M13]{NGC~6205 (M13)}}
\newcommand{\ngcc}{\object{M13}}
\newcommand{\ngcdf}{\object[M71]{NGC~6838 (M71)}}
\newcommand{\ngcd}{\object{M71}}
\newcommand{\ngce}{\object{NGC 7006}}
\newcommand{\ngcff}{\object[M79]{NGC~7078 (M15)}}
\newcommand{\ngcf}{\object{M15}}
\newcommand{\ngcg}{\object{NGC 7492}}
\newcommand{\leoi}{\object[UGC 5470]{Leo~I}}
\newcommand{\andromeda}{\object{M31}}
\newcommand{\bd}{\object[TYC 2214- 1198-1]{BD+28$^{\circ}$~4211}}
\newcommand{\arcturus}{\object[NAME ARCTURUS]{Arcturus}}

\title{Metallicity and Alpha-Element Abundance Measurement in Red
  Giant Stars from Medium Resolution Spectra}

\author{Evan~N.~Kirby, Puragra~Guhathakurta}
\affil{University of California Observatories/Lick Observatory,
  Department of Astronomy \& Astrophysics, \\
  University of California, Santa Cruz, CA 95064; \email{ekirby@ucolick.org}}
\and
\author{Christopher Sneden}
\affil{Department of Astronomy and McDonald Observatory \\
  University of Texas, 1 University Station, C1400, Austin, TX 78712}


\begin{abstract}
We present a technique that applies spectral synthesis to medium
resolution spectroscopy (MRS, $R \sim 6000$) in the red
($6300~\mathrm{\AA} < \lambda < 9100$~\AA) to measure [Fe/H] and
[$\alpha$/Fe] of individual red giant stars over a wide metallicity
range.  We apply our technique to 264 red giant stars in seven
Galactic globular clusters and demonstrate that it reproduces the
metallicities and $\alpha$ enhancements derived from high resolution
spectroscopy (HRS).  The MRS technique excludes the three \ion{Ca}{2}
triplet lines and instead relies on a plethora of weaker lines.
Unlike empirical metallicity estimators, such as the equivalent width
of the \ion{Ca}{2} triplet, the synthetic method presented here is
applicable over an arbitrarily wide metallicity range and is
independent of assumptions about the $\alpha$ enhancement.  Estimates
of cluster mean [Fe/H] from different HRS studies show typical scatter
of $\sim 0.1$~dex but can be larger than 0.2~dex for metal-rich
clusters.  The scatter in HRS abundance estimates among individual
stars in a given cluster is also comparable to 0.1~dex.  By
comparison, the scatter among MRS [Fe/H] estimates of individual stars
in a given cluster is $\sim 0.1$~dex for most clusters but 0.17~dex
for the most metal-rich cluster, \object{M71}
($\langle\mathrm{[Fe/H]}\rangle = -0.8$).  A star-by-star comparison
of HRS vs.\ MRS [$\alpha$/Fe] estimates indicates that the precision
in $\mathrm{[\alpha/Fe]}_{\mathrm{MRS}}$ is 0.05~dex.  The errors in
$\mathrm{[Fe/H]}_{\mathrm{MRS}}$ and
$\mathrm{[\alpha/Fe]}_{\mathrm{MRS}}$ increase beyond 0.25~dex only
below signal-to-noise ratios of $20~\mathrm{\AA}^{-1}$, which is
typical for existing MRS of the red giant stars in \object[UGC
5470]{Leo~I}, one of the most distant Milky Way satellites
(250~kpc).

\end{abstract}



\section{Introduction}
\label{sec:intro}
In the popularly accepted paradigm of hierarchical structure formation
\citep{sea78,whi78}, large galaxies grow by assembling smaller
components.  Simulations motivated by $\Lambda$CDM cosmology have
successfully reproduced the detailed properties of observed galaxies
\citep[e.g.,][]{rob05, dut05}.  Such simulations have advanced enough
to track an important prediction of $\Lambda$CDM cosmology: the
chemical properties of stars in different dynamical components in a
single dark matter halo, including the cold dwarf satellites, the
dissolving tidal streams, and the hot stellar halo.  For example,
\citet{fon06} make predictions of metallicity and $\alpha$
enhancements of a Milky Way-like halo.  They verify their metallicity
predictions with photometric measurements of \andromeda\
\citep{fon08}.

Several techniques can be used to measure stellar abundances.  The
most trusted abundance measurement tool is high resolution
spectroscopy (HRS).  A high resolution stellar spectrum contains
information about the temperature, surface gravity, and abundances of
individual elements in the stellar atmosphere.  However, HRS spreads
the light of a star over finely spaced resolution elements and
therefore requires long exposures of bright stars.  With the current
state-of-the-art telescopes, typical HRS targets are brighter than $V
\sim 15$, although some authors have observed targets as faint as $V
\sim 18$ \citep[e.g.,][]{coh05a}.  HRS targets have ranged from
moderate to low luminosity stars in the solar neighborhood to high
luminosity stars throughout the Milky Way (MW) and even in its dwarf
satellites \citep[e.g.,][]{she03}.  HRS within the MW has allowed
measurements of \feh, \afe, and individual element abundances of
individual stars in different dynamical components
\citep[e.g.,][]{ven04}.  However, it is currently not feasible to
obtain large HRS samples of individual stars beyond the MW and its
satellites.

In order to reach fainter and more distant stellar systems, photometry
of resolved stellar populations is a commonly used abundance
measurement tool.  Most studies based on color-magnitude diagrams
(CMDs) target the most luminous components of a population---the red
giant branch (RGB) and horizontal branch (HB)---because they are
visible over large distances.  CMDs that do not reach the main
sequence turn-off (MSTO) are susceptible to the age-metallicity
degeneracy and the HB second parameter problem \citep{san67}.  As a
rare exception, \citet{bro03} and \citet{bro06} have used extremely
deep imaging of three fields in \andromeda\ to constrain not only
metallicity distributions but also ages of stellar populations.
However, such deep photometry of \andromeda\ is extremely expensive
and infeasible for more than a few fields.  Even with the MSTO,
photometry provides only one abundance dimension, overall metallicity.
In fact, photometric abundance estimates based on the RGB must assume
\afe.

Medium resolution spectroscopy (MRS) avoids most of the assumptions
involved in photometric metallicity estimates.  Furthermore, it can
provide multiple abundance dimensions, such as \feh\ and \afe.
State-of-the-art MRS instruments include \textsc{lris} and \deimos\ on
the Keck telescopes, \textsc{fors} and \textsc{vimos} on the Very
Large Telescope, and \textsc{imacs} on Magellan.  Some of these
instruments have been used to obtain MRS of RGB stars in the bulge,
disk, halo, and dwarf satellites of \andromeda\
\citep{guh06,cha06,kal06,koc07}.

The traditional MRS abundance technique is spectrophotometric indices.
\citet{pre61} established the first calibration between the equivalent
width (EW) of a single metal line (\ion{Ca}{2}~K $\lambda 3933$) and
stellar metallicity.  The \ion{Ca}{2} $\lambda \lambda \lambda
8498,8542,8662$ infrared triplet has been a more popular metallicity
indicator in the last decade.  \citet{arm88} first quantified the
correlation between the EW of the \ion{Ca}{2} triplet in integrated
light spectra of MW GCs and their previously measured \feh.  A
plethora of astronomers have since developed their own \ion{Ca}{2}
triplet metallicity calibrations, but \citet{ols91} and \citet{rut97}
are the most highly cited.  Despite the success of the \ion{Ca}{2}
triplet as a metallicity indicator over a wide range ($-2.2 \lesssim
\mathfeh \lesssim -0.6$), its use in determining \feh\ necessitates
assuming the ratio [Ca/Fe].  In fact, \citeauthor{rut97} warn,
``caution---perhaps considerable---may be advisable when using
[\ion{Ca}{2} triplet reduced width] as a surrogate for metallicity,
especially for systems where ranges in age and metallicity are
likely.''  For an analysis of the different effects of \feh\ and
[Ca/H] on the \ion{Ca}{2} triplet EW, see \citet{bat08}, who also
claim errors of 0.1~dex in \feh\ at signal-to-noise ratios
$\mathrm{S/N} > 10~\mathaa^{-1}$.  Finally, any empirical calibration
is restricted to the metallicity range of the calibrators.  The
\ion{Ca}{2} triplet metallicities may not be accurate below $\mathfeh
< -2.2$ \citep{koc08}.

Spectral modeling, essentially the HRS abundance technique,
circumvents dependence on the properties of calibrators.  The
procedure is to synthesize a spectrum of a stellar atmosphere and
compare to an observed spectrum.  The atmospheric parameters and
abundances can be adjusted until the best fit is achieved.  Most HRS
studies model individual lines and compare synthetic and observed EWs.
At medium and low resolution, most metal absorption lines are weak,
blended, or both.  However, a few thousand \AA\ of spectral coverage
includes hundreds of metal absorption lines.  Pixel-to-pixel spectral
fits leverage the statistical power of many lines to find the optimal
atmosphere and abundance.

This method is not new to MRS.  For example, \citet{sun81} and
\citet{car82} modeled blue spectra at low resolution to determine C
and N abundances in globular cluster (GC) RGB stars.  Synthetic
pixel-to-pixel matching is one metallicity estimator in the Sloan
Extension for Galactic Understanding and Exploration (SEGUE) Stellar
Parameter Pipeline \citep{lee07,all06}.  Future versions of the
pipeline will even estimate \afe.  The unique feature of our work is
the use of only red and far-red spectral regions ($6300~\mathaa <
\lambda < 9100$~\AA).  We have chosen this spectral region to make use
of existing spectra of stars in the dwarf satellites of the MW and in
the stellar halo and dwarf satellites of \andromeda.

As a general cautionary note on spectroscopic abundance estimates,
studies near the tip of the RGB are sensitive to modeling assumptions,
including local thermodynamic equilibrium, atmospheric geometry, and
the mode of energy transport.  However, these assumptions are
significantly less severe than those involved in photometric
metallicity measurements, and they become rapidly less significant at
lower luminosities.

To assess the accuracy and precision of the MRS technique, we compare
our results to measurements of the same stars in the GCs observed with
HRS.  We discuss our observations in \S\,\ref{sec:obsdata} and the
preparation of the spectra for abundance analysis in
\S\,\ref{sec:analysis}.  The abundance measurements are described in
\S\,\ref{sec:abundanalysis}.  We compare the MRS and HRS results in
\S\,\ref{sec:results}, and we quantify errors in
$\mathfeh_{\mathrm{MRS}}$ and $\mathafe_{\mathrm{MRS}}$ in
\S\,\ref{sec:error}.  We discuss the range of applications for the MRS
technique, the ways it can improve, and its potential advantages over
other medium resolution techniques in \S\,\ref{sec:applications}.


\section{Observations and Data Reduction}
\label{sec:obsdata}
\subsection{Observations}
\label{sec:observations}

\begin{deluxetable*}{lcccc}
\tablecolumns{5} \tablecaption{Observations\label{tab:observations}}
\tablehead{\colhead{Object} & \colhead{Date} & \colhead{Airmass} &
\colhead{Exposures} & \colhead{\# targets}} \startdata
\ngcaf\tablenotemark{a} & 2006 Feb 2\phn & 1.42 & 2 $\times$ 300~s & \phn22 \\
\ngcb\tablenotemark{a} & 2006 Feb 2\phn & 1.21 & 4 $\times$ 300~s & \phn70 \\
\ngccf & 2007 Oct 12 & 1.35 & 3 $\times$ 300~s & \phn93 \\
\ngcdf & 2007 Nov 13 & 1.09 & 3 $\times$ 300~s & 104 \\
\ngce & 2007 Nov 15 & 1.01 & 2 $\times$ 300~s & 105 \\
\ngcff & 2007 Nov 14 & 1.01 & 2 $\times$ 300~s & \phn63 \\
\ngcg & 2007 Nov 15 & 1.30 & 2 $\times$ 210~s & \phn38 \\
\enddata
\tablenotetext{a}{\citet{sim07} have generously provided these
observations.}
\end{deluxetable*}

We present observations of seven GCs with the Deep Imaging
Multi-Object Spectrometer \citep[\deimos,][]{fab03} on the Keck
\textsc{ii} telescope.  Table~\ref{tab:observations} lists the
clusters observed, the dates of observations, the exposure times, and
the number of stars targeted.  Because the targets are very bright, we
observed near $12\arcdeg$ dusk twilight without focusing the primary
mirror.  Subsequent focusing resulted in insignificant mirror segment
alignment.  The seeing was very good, near $0.5''$ for most
observations.  We used the OG550 order-blocking filter with the 1200
lines~mm$^{-1}$ grating and $0.7''$ slit widths.  This configuration
mimics the \deimos\ configurations of \andromeda\ RGB stars
\citep{guh06} and RGB stars in \leoi, a remote dSph satellite of the
MW \citep{soh07}.  The spectral resolution is $\sim 1.3$~\AA\ FWHM
(resolving power $R \sim 6000$).  The spectral range is about
6300--9100~\AA\ with variation depending on the slit's location along
the dispersion axis.  Exposures of Kr, Ne, Ar, and Xe arc lamps
provided wavelength calibration, and exposures of a quartz lamp
provided flat fielding.

One $16' \times 4'$ slitmask was devoted to each cluster.  Each slit
included one star.  We attempted to maximize the number of target
stars previously observed with HRS.  (The slitmask for \ngca\ is an
exception, and it contains no targets previously observed with HRS.)
We selected the remaining targets based on CMDs.  In order of
priority, we filled each slitmask with stars from the (1) upper RGB,
(2) lower RGB, (3) red clump, and (4) blue HB.  We filled the slits at
the edges of the slitmask far from the center of the GCs with objects
with similar colors and magnitudes to stars on the RGB.

\subsection{Data reduction}
\label{sec:reduction}
We reduce the raw frames using version 1.1.4 of the \deimos\ data
reduction pipeline developed by the \deep\ Galaxy Redshift
Survey.\footnote{\url{http://astro.berkeley.edu/$^{\sim}$cooper/deep/spec2d/}}
\citet{guh06} give the details of the data reduction.  We also make
use of the optimizations to the code described in \citet[\S\,2.2 of
their article]{sim07}.  These modifications provide better extraction
of unresolved stellar sources.

In summary, the pipeline traces the edges of slits in the flat field
to determine the CCD location of each slit.  The wavelength solution
is given by a polynomial fit to the CCD pixel locations of arc lamp
lines.  Each exposure of stellar targets is rectified and then
sky-subtracted based on a B-spline model of the night sky emission
lines.  Next, the exposures are combined with cosmic ray rejection
into one two-dimensional spectrum.  Finally, the one-dimensional
stellar spectrum is extracted from a small spatial window in the
two-dimensional spectrum encompassing the light of the star.  The
product of the pipeline is a wavelength-calibrated, sky-subtracted,
cosmic ray-cleaned, one-dimensional spectrum for each target.


\section{Spectral Analysis}
\label{sec:analysis}
\subsection{Determination of spectral resolution}
When we compare model spectra to observed spectra, we must match the
synthetic and observed resolving power $R$.  For our configuration of
\deimos, $R$ is a slight function of wavelength, typically varying
from 5,500 to 7,200.  We determine the observed spectrum's resolution
by fitting Gaussians to hundreds of sky lines in the same slit as the
object spectrum.  Then, we fit a parabola to the Gaussian widths as a
function of wavelength.  For some short slits, this procedure fails.
In those cases, we simply fit a parabola to the measured Gaussian
widths of sky lines from all slits on the same \deimos\ slitmask as a
function of observed wavelength.

\subsection{Telluric absorption correction}
The last step in obtaining a stellar-only spectrum is removal of
terrestrial atmospheric absorption.  We build a telluric absorption
template from a continuum-divided spectrum of a hot star free of metal
absorption lines.  On 2007 November 14, we observed the white dwarf
spectrophotometric standard \bd\ with a longslit in the same
spectrometric configuration as the slitmasks.  The airmass was 1.018.
We assume that all detectable absorption lines in the spectrum except
\ha\ are telluric.

The spectral regions most susceptible to telluric absorption are
$6864~\mathaa < \lambda < 7020$~\AA\ (B band), $7162~\mathaa < \lambda
< 7350$~\AA, $7591~\mathaa < \lambda < 7703$~\AA\ (A band),
$8128~\mathaa < \lambda < 8352$~\AA, and $\lambda > 8938$~\AA.  In
order to normalize the continuum of the telluric absorption template,
we simply fit a line to the 100~\AA\ bands on either side of each
region.  Then, we divide each region by its best-fit line.  Because
the hot star shows no detectable telluric absorption outside of these
regions, we set all remaining pixels to 1 to prevent introducing noise
during telluric absorption removal.

We interpolate the telluric absorption template onto the wavelength
array for each star observed.  Then, we divide the observed spectrum
by the template adjusted by the ratio of the airmasses, following the
Beer-Lambert law:

\begin{equation}
d = \frac{s}{t^{X(\mathrm{obs}) / X(\mathrm{tell})}}
\end{equation}

\noindent
where $X$ is the airmass, $d$ is the telluric-corrected spectrum, $s$
is the raw spectrum, and $t$ is the telluric absorption template.  We
carefully adjust the observed spectrum's variance array, treating the
noise in the telluric spectrum as uncorrelated with the noise in the
observed spectrum.

The amount of attenuation varies as a function of the number density
of absorbers---particularly water vapor---along the line of sight.
The associated frequency-dependent optical depth varies on timescales
possibly less than an hour.  Regardless, the telluric absorption
removal procedure described here works very well even for observations
taken 21 months before the telluric standard.  The procedure does a
poor job in three spectral regions: the saturated A band, the
saturated portion of the B band, and a small 40~\AA\ region around
8245~\AA.  We ignore these regions in the abundance analysis.

\begin{figure}
\plotone{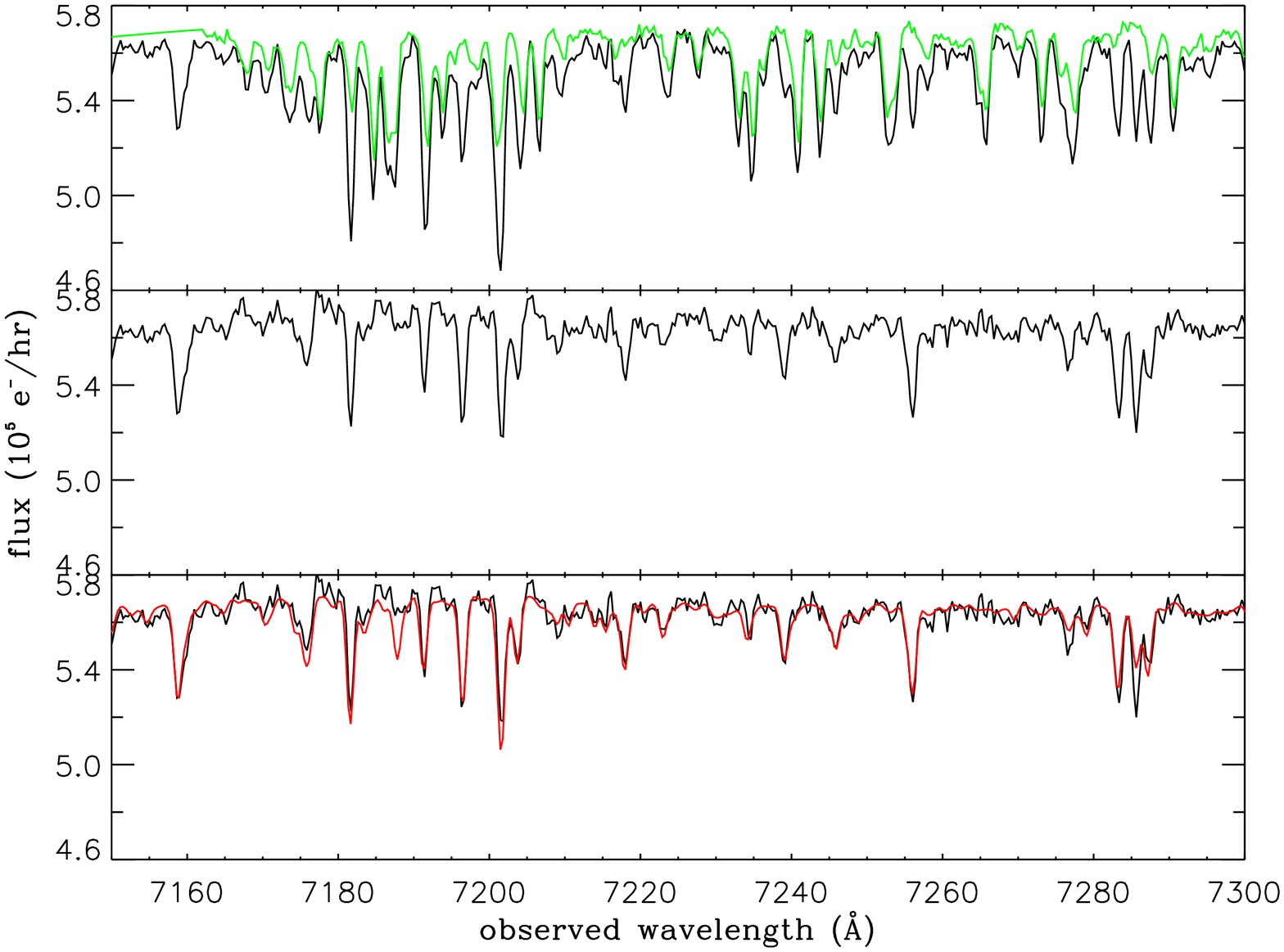}
\caption{A section of the spectrum for one star in M13.  {\it Top:}
  The observed spectrum (black) with the telluric absorption template
  (green).  Every feature in the telluric absorption template has a
  counterpart in the observed spectrum, but the reverse is not true.
  {\it Middle:} Applying the telluric absorption correction completely
  removes some of the absorption features seen in the raw spectrum,
  makes a few of the absorption features weaker, and leaves many of
  the features unaffected.  {\it Bottom:} The telluric-divided
  spectrum (black, same as in middle panel) with the best-fit
  synthetic spectrum (red).  The close agreement between the two
  spectra demonstrates that the features remaining in the observed
  spectrum are intrinsic to the star.  Both the green and red lines
  have been scaled by the measured continuum.\label{fig:telluric}}
\end{figure}

Figure~\ref{fig:telluric} demonstrates the efficacy of the telluric
absorption removal.  The upper panel shows part of an example spectrum
before the telluric correction, along with the telluric absorption
template.  The middle panel shows the example spectrum after the
correction.  The bottom panel includes a synthetic stellar spectrum
(see details in \S\,\ref{sec:analysis}), which shows that every
absorption line left after the telluric correction is intrinsic to the
star.

\subsection{Radial velocities}
\label{sec:velocities}
We measure radial velocities of each target star to check its cluster
membership and to shift its spectrum to the rest frame for abundance
measurement.  We cross-correlate each telluric-divided observed
spectrum with a synthetic spectrum (see details in
\S\,\ref{sec:abundanalysis}) with $\mathteff = 4500$~K, $\mathlogg =
1.5$, $\mathfeh = -1.5$, and $\mathafe = +0.2$ in the spectral region
$8450~\mathaa < \lambda < 8700~\mathaa$.  This region includes the
\ion{Ca}{2} triplet, which is strong even in hot, extremely metal-poor
stars, making it ideal for radial velocity determination in a wide
range of stars.  We shift the observed spectra to the rest frame to
complete the remainder of the analysis.

\subsection{Continuum determination}
\label{sec:continuum}
The abundance measurements are particularly sensitive to an accurate
determination of the continuum.  Underestimating the continuum will
make absorption lines appear too shallow, and the derived abundances
will be too low, the temperature too high, or both.  The abundance
analysis described in \S\,\ref{sec:abundanalysis} is insensitive to
the global continuum shape and instead relies on high-frequency
line-to-line variations in flux as a function of wavelength.
Therefore, we have decided to focus on local continuum determination.

First, we determine the spectral regions free of stellar absorption.
Following the procedures in \S\,\ref{sec:abundanalysis}, we generate a
synthetic spectrum between 6300~\AA\ and 9100~\AA\ of a star with
$\mathteff = 4300$~K, $\mathlogg = 1.5$, and $\mathfeh = -0.5$.  The
spectrum is smoothed through a moving Gaussian kernel of $\sigma =
0.6$~\AA\ to simulate the approximate spectral resolution of \deimos.
The synthetic spectrum has a perfectly flat continuum, and the units
are such that the continuum is 1.  We call spectral regions with
synthetic flux greater than 0.96 and a minimum width of 0.5~\AA\
``continuum regions.''  Pixels at observed wavelengths outside of
these regions will not contribute to the continuum determination of
observed spectra.

Next, we compute the continuum.  Each pixel in the continuum ($c_j$)
is the weighted average of its neighboring pixels in the observed
spectrum ($s_i$).  The weight is a combination of the inverse variance
($\sigma_i^2$) and proximity in wavelength.

\begin{figure}
\plotone{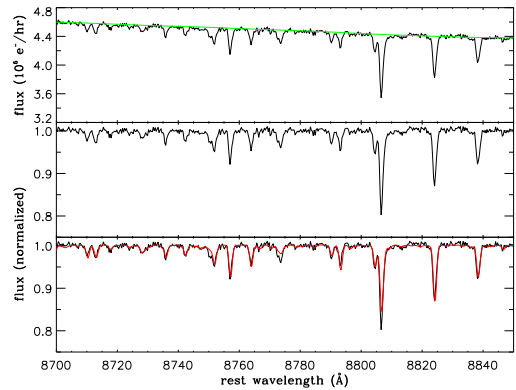}
\caption{A different section of the same star as in
  Fig.~\ref{fig:telluric}.  {\it Top:} The observed spectrum (black)
  and the measured continuum (green).  {\it Middle:} The observed
  spectrum divided by the continuum.  {\it Bottom:} The
  continuum-normalized observed spectrum (black) with the best-fit
  synthetic spectrum (red), which has also been subjected to continuum
  normalization (see \S\,\ref{sec:continuum} and
  \S\,\ref{sec:abundance}).\label{fig:continuum}}
\end{figure}

\begin{eqnarray}
c_j &=& \frac{\sum_i s_i w_{i,j} m_i}{\sum_i w_{i,j} m_i} \\
w_{i,j} &=& \frac{1}{\sigma_i^2}\exp{\left(-\frac{(\lambda_i - \lambda_j)^2}{2 (10.0~\mathaa)^2}\right)} \\
m_i &=& \left\{
\begin{array}{rl}
1 & \mathrm{if~} \lambda_i \in \mathrm{continuum~region} \\
0 & \mathrm{otherwise}
\end{array} \right.
\end{eqnarray}

\noindent
This process simultaneously accomplishes smoothing and interpolation
across non-continuum regions.  We normalize each observed spectrum by
dividing by its measured continuum.  Figure~\ref{fig:continuum} shows
the measured continuum in one example spectrum.

The continuum regions include some weak absorption lines, which will
drag the continuum determination down by up to 4\% in the coolest,
most metal rich stars in our sample.  In \S\,\ref{sec:abundance}, we
determine the synthetic spectrum that most closely matches the
observed spectrum.  Before comparing a synthetic spectrum to the
observed one, we apply the above continuum determination technique to
both.  Even though the synthetic spectrum has a perfectly flat
continuum, application of this continuum determination technique
causes the same degree of weak feature continuum suppression in both
the synthetic and observed spectra, thereby correcting this effect.

\subsection{Pixel mask}
\label{sec:mask}

\begin{deluxetable*}{lcc}
\tablecolumns{3}
\tablecaption{Spectral Masks\label{tab:mask}}
\tablehead{\colhead{Feature}  &  \colhead{Lower Wavelength (\AA)}  &  \colhead{Upper Wavelength (\AA)}}
\startdata
\cutinhead{Telluric Absorption}
B band  				  &  6864\phd\phn\phn\phn  &  7020\phd\phn\phn\phn  \\
A band  				  &  7591\phd\phn\phn\phn  &  7703\phd\phn\phn\phn  \\
strong telluric absorption                &  8225\phd\phn\phn\phn  &  8265\phd\phn\phn\phn  \\
\cutinhead{Stellar Absorption}
\ion{Ca}{1} $\lambda$6343         	  &  6341\phd\phn\phn\phn  &  6346\phd\phn\phn\phn  \\
\ion{Ca}{1} $\lambda$6362  	      	  &  6356\phd\phn\phn\phn  &  6365\phd\phn\phn\phn  \\
\ha                                       &  6559.797              &  6565.797  \\
\ion{K}{1} $\lambda$7665   	      	  &  7662\phd\phn\phn\phn  &  7668\phd\phn\phn\phn  \\
\ion{V}{1} $\lambda$8116,$\lambda$8119 hyperfine structure  &  8113\phd\phn\phn\phn  &  8123\phd\phn\phn\phn  \\
poorly modeled absorption in \arcturus    &  8317\phd\phn\phn\phn  &  8330\phd\phn\phn\phn  \\
\ion{Ca}{2} $\lambda$8498  		  &  8488.023  		   &  8508.023  \\
\ion{Ca}{2} $\lambda$8542  		  &  8525.091  		   &  8561.091  \\
\ion{Ca}{2} $\lambda$8662  	          &  8645.141  		   &  8679.141  \\
\ion{Mg}{1} $\lambda$8807  		  &  8804.756  		   &  8809.756  \\
\enddata
\end{deluxetable*}

Each spectrum has its own pixel mask.  Table~\ref{tab:mask} lists
spectral regions in observed and rest wavelength that may confuse the
abundance analysis.  The first part of the table lists regions of
saturated or very strong telluric absorption.  We mask the pixels that
fall in these regions before we shift the spectrum to zero velocity.
The second part of the table lists stellar absorption features that
are difficult to model.  We mask the pixels that fall in these regions
after shifting the spectrum to zero velocity.  These features include
the strongest $\alpha$ element lines in the spectral range of \deimos:
the \ion{Ca}{2} triplet and \ion{Mg}{1} $\lambda$8807.  The cores of
the \ion{Ca}{2} lines are formed very high in the stellar atmosphere
out of local thermodynamic equilibrium (LTE).  Because we could not
synthetically reproduce the width and shape of the \ion{Mg}{1} line in
Arcturus (see \S\,\ref{sec:linelist}), we excluded it from the
abundance analysis.

In addition to the standard mask in Table~\ref{tab:mask}, we inspected
each spectrum to mask improperly subtracted sky lines, cosmic rays,
and other obvious instrumental artifacts.


\section{Chemical Abundance Analysis}
\label{sec:abundanalysis}

\begin{deluxetable*}{lcccr}
\tablecolumns{5}
\tablewidth{0pc}
\tablecaption{Grid of Synthetic Spectra\label{tab:grid}}
\tablehead{\colhead{Parameter}  &  \colhead{Lower Limit}  &  \colhead{Upper Limit}  &  \colhead{Step}  &  \colhead{Number}}
\startdata
\teff~(K)           &  4000     &     8000  &  $\left\{\begin{array}{rl}
                                               100 & \mathrm{if~} \mathteff \le 5500 \\
                                               200 & \mathrm{if~} \mathteff \ge 5600
                                               \end{array}\right.$  & 29 \\
\logg               &  $\left\{\begin{array}{r}
                       0.0 \\
                       0.0 \\
                       0.5 \\
                       0.5 \\
                       1.0 \\
                       0.5 \\
                       1.0 \\
                       1.0 \\
                       1.5
                       \end{array}\right.$  &
                       $\begin{array}{rl}
                       4.5 & \mathrm{if~} \mathteff < 4500 \mathrm{~and~} \mathfeh < -2.5 \\
                       5.0 & \mathrm{if~} (\mathteff < 4500 \mathrm{~and~} \mathfeh \ge -2.5) \mathrm{~or~} 4500 \le \mathteff \le 6000 \\
                       5.0 & \mathrm{if~} 6000 < \mathteff \le 7000 \\
                       5.0 & \mathrm{if~} 7000 < \mathteff \le 7200 \mathrm{~and~} \mathfeh \ge -3.5 \\
                       5.0 & \mathrm{if~} 7000 < \mathteff \le 7200 \mathrm{~and~} \mathfeh < -3.5 \\
                       5.0 & \mathrm{if~} 7200 < \mathteff \le 7500 \mathrm{~and~} \mathfeh \ge -2.5 \\
                       5.0 & \mathrm{if~} 7200 < \mathteff \le 7500 \mathrm{~and~} \mathfeh < -2.5 \\
                       5.0 & \mathrm{if~} \mathteff > 7500 \mathrm{~and~} \mathfeh \ge -2.5 \\
                       5.0 & \mathrm{if~} \mathteff > 7500 \mathrm{~and~} \mathfeh < -2.5
                       \end{array}$  &  0.5  &  $\left\{\begin{array}{r}
                                                10 \\
                                                11 \\
                                                10 \\
                                                10 \\
                                                 9 \\
                                                10 \\
                                                 9 \\
                                                 9 \\
                                                 8
                                                \end{array}\right.$ \\
\feh                &   $-4.0$  &  \phs0.0  &  0.1  &       41 \\
\afe                &   $-0.6$  &   $+1.0$  &  0.1  &       17 \\
\tableline
Total               &           &           &       &  210 902 \\
\enddata
\end{deluxetable*}

The heart of this analysis is a very large grid of synthetic stellar
spectra.  The grid contains four dimensions: effective temperature
(\teff), surface gravity (\logg), metallicity (\feh), and alpha
enhancement (\afe).  Table~\ref{tab:grid} lists the details of the
grid.

In this article, we use the standard spectroscopic abundance notation.
The ratio of any two elements A and B in a star relative to their
ratio in the Sun is

\begin{equation}
\mathrm{[A/B]} \equiv \log [n(\mathrm{A})/n(\mathrm{B})] - \log
[n(\mathrm{A})/n(\mathrm{B})]_{\sun}
\end{equation}

\noindent
where $n$ is number density.\footnote{In this article, $12 + \log
[n(\mathrm{Fe})/n(\mathrm{H})]_{\sun} = 7.52$ \citep[as adopted
by][]{sne92}.  The abundances of all other elements are the solar
values of \citet{and89} except Li, Be, and B, for which we use the
meteoritic values.}  \feh\ represents metallicity, the overall bulk
content of the heavy elements.  Because neutral Fe and neutral
$\alpha$ element lines dominate the red spectra of stars in our
\teff-\logg-\feh\ domain, we assume $\mathrm{[X/H]} = \mathfeh$ for
most elements.  We stress, however, that lines of all elements except
the $\alpha$ elements contribute to the MRS measurement of \feh.  The
abundances of the $\alpha$ elements Mg, Si, S, Ar, Ca, and Ti are
modified by the additional parameter \afe:

\begin{equation}
\mathrm{[\alpha/H]} = \mathrm{[Fe/H]} + \mathrm{[\alpha/Fe]}\:.
\end{equation}

\noindent
The abundance enhancements of all six elements vary together.

We have found the best agreement with HRS by fixing the microturbulent
velocity (\vt) to \logg\ with an empirical relation.  The best linear
fit to the RGB sample ($\mathlogg < 3.3$) of \citet{ful00} is

\begin{equation}
v_t = 2.700 - 0.509\,\log g~\mathrm{km/s}.
\end{equation}

\subsection{Model atmospheres}
\label{sec:atmospheres}
The starting point for generating a synthetic spectrum is a model
stellar atmosphere, which is a tabulation of temperature, pressure,
electron fraction, and opacity as a function of optical depth.  We
choose to use the Castelli \& Kurucz grid of models with no convective
overshooting \citep{cas97}.  For atmospheres with $\mathfeh \ge
-2.5$, we use the ``ODFNEW'' models, with updated opacity distribution
functions \citep{cas04}.

The published model atmosphere grid points of \teff, \feh, and \afe\
are coarser than the step sizes listed in Table~\ref{tab:grid}.
Therefore, for a fixed array of optical depths, we linearly
interpolate temperatures, pressures, electron fractions, and opacities
between model atmosphere grid points to generate a single model
atmosphere for every grid point in Table~\ref{tab:grid}.

The alpha enhancement changes the ionization balance and free election
fraction in the atmosphere.  ODFNEW ($\mathfeh \ge -2.5$) models are
available for $\mathafe = 0.0$ and $+0.4$.  For synthesizing spectra
with $\mathafe \le 0.0$, we choose the former and for spectra with
$\mathafe \ge +0.4$, we choose the latter.  For intermediate values of
$\mathafe$, we linearly interpolate between the two regimes.
Atmospheres with $\mathfeh < -2.5$ are available only with $\mathafe =
0.0$.  For the model atmospheres only, the $\alpha$ elements are O,
Ne, Mg, Si, S, Ar, Ca, and Ti.

\subsection{Line lists}
\label{sec:linelist}
We assembled a line list of wavelengths, excitation potentials (EP),
and oscillator strengths ($gf$) for atomic and molecular transitions
that occur in the red spectral regions of stars in our stellar
parameter range.  In standard HRS analyses, it is often possible to
use only transitions with accurately measured oscillator strengths.
For this MRS analysis that covers a broad spectral range with many
blended lines, we must use a database of lines with oscillator
strengths of varying accuracy.

To begin, we queried the Vienna Atomic Line Database
\citep[VALD,][]{kup99} for all transitions of neutral or singly
ionized atoms with $\mathrm{EP} < 10$~eV and $\log gf > -5$.  We
supplemented the list with CN, $\mathrm{C}_2$, and MgH molecular
transitions \citep{kur92} and Li, Sc, V, Mn, Co, Cu, and Eu hyperfine
transitions \citep{kur93}.  By far, CN is the most important molecule
for our red spectra.  TiO is also strongly present in cool, metal-rich
stars.  However, the TiO system is extremely complex and difficult to
model accurately.  The large number of TiO electronic transitions in
red spectra makes TiO spectral synthesis computationally daunting.
Fortunately, TiO becomes a significant absorber only in metal-rich
stars with $\mathteff < 4000$~K.  Therefore, we did not include TiO in
our line list, and we chose $\mathteff = 4000$~K as the grid's lower
limit.

Next, we generated synthetic spectra of the Sun and \arcturus\ using
our line list, model atmospheres as described in
\S\,\ref{sec:atmospheres}, and the current version of the LTE spectral
synthesis software MOOG \citep{sne73}.  The spectral range is
6300~\AA\ to 9100~\AA, and the resolution is 0.02~\AA.  The line
broadening accounts for collisions with neutral hydrogen for for lines
with tabulated damping constants \citep{bar00,bar05}.  For other
lines, the Uns{\"o}ld approximation multiplied by 6.3 gives the van
der Waals line damping parameter, and MOOG calculates additional
radiative and Stark broadening.  However, the choice of damping
parameter does not noticeably affect the spectra of low-pressure giant
stars such as Arcturus, nor any star in the domain of \teff\ and
\logg\ that we consider here.

The atmospheric parameters for the Sun are $\mathteff = 5798$~K,
$\mathlogg = 4.44$, and $\mathfeh = 0.0$.  For \arcturus, we adopt the
atmospheric parameters and non-solar abundance ratios determined by
\citet{pet93}: $\mathteff = 4300$~K, $\mathlogg = 1.50$, and
$\mathfeh = -0.50$.\footnote{\citet{pet93} use a value of the solar
iron abundance that is 0.15~dex higher than the one adopted in this
article.  In fact, they note that decreasing the iron abundance to the
value used here reproduces the profiles of weak \ion{Fe}{1} lines
without adjusting laboratory-measured oscillator strengths.  We find
that their published value of \feh\ with our value of the solar iron
abundance reproduces the spectrum of \arcturus\ very well.}  We
compared these two synthetic spectra to the \citet{hin00} atlases of
the Sun and \arcturus.  After smoothing the synthetic spectra through
a Gaussian kernel to match the line profiles of the atlas, we
inspected the spectra in detail.  First, we adjusted oscillator
strengths of aberrant atomic lines in the solar synthetic spectrum
until the strengths of the synthesized lines matched that of the
observed lines.  Then, we repeated the process for \arcturus, making
adjustments that did not cause disagreement in the solar spectrum.

\begin{deluxetable}{lccc}
\tablecolumns{4}
\tablecaption{Line List\label{tab:linelist}}
\tablehead{\colhead{Wavelength (\AA)}  &  \colhead{Species}  &  \colhead{EP (eV)}  &  \colhead{$\log gf$}}
\startdata
8647.703  &  \phn22.0\phn\phn  &  4.654        &     $-3.561$        \\
8647.792  &  \phn25.1\phn\phn  &  6.834        &     $-2.489$	     \\
8647.799  &  \phn21.0\phn\phn  &  4.558        &     $-1.759$	     \\
8647.807  &  \phn26.0\phn\phn  &  5.720        &     $-1.519$	     \\
8647.940  &  \phn26.0\phn\phn  &  6.5\phn\phn  &     $-0.5$\phn\phn  \\
8648.024  &  \phn21.0\phn\phn  &  4.126        &     $-2.710$	     \\
8648.040  &     607.0\phn\phn  &  1.093        &     $-3.058$        \\
8648.366  &     607.0\phn\phn  &  1.093        &     $-1.096$        \\
8648.400  &     607.0\phn\phn  &  0.912        &     $-2.763$        \\
8648.455  &  \phn14.0\phn\phn  &  6.21\phn     &  \phs$0.0$\phn\phn  \\
8648.556  &  \phn16.0\phn\phn  &  8.408        &     $-1.720$	     \\
8648.630  &  \phn27.059        &  2.280        &     $-3.852$	     \\
8648.688  &     607.0\phn\phn  &  0.912        &     $-1.466$        \\
8648.699  &  \phn27.059        &  2.280        &     $-4.068$	     \\
8648.716  &  \phn20.0\phn\phn  &  4.554        &     $-2.024$        \\
\enddata
\tablecomments{Table~\ref{tab:linelist} is published in its entirety
  in the electronic edition of the Astrophysical Journal.  A portion
  is shown here for guidance regarding form and content.  The first
  column is the line wavelength.  The second is a code representing
  the atomic or molecular species.  The integer part of the code is
  the element's atomic number.  Codes greater than 100 represent
  molecules.  For example, 607 represents CN.  The first decimal place
  of the code is the ionization state, where 0 is neutral.  Remaining
  decimal places are mass numbers for isotopic hyperfine transitions.
  The third column is EP.  The invented \ion{Fe}{1} transitions have
  EP with fewer than three decimal places.  The fourth column is
  oscillator strength.  The transitions that we modified have
  oscillator strengths with fewer than three decimal places.}
\end{deluxetable}

Occasionally, we encountered observed lines not present in the line
list.  Most of these cases could be resolved by relaxing the $\log gf
> -5$ restriction.  However, in cases where a single line was not
represented in the VALD or Kurucz line lists, we invented a transition
of \ion{Fe}{1} of the EP and $gf$ required to reproduce the line's
observed strength in both the Sun and \arcturus.  The final line list
contains 30873 atomic and 17345 molecular transitions, representing 71
elements.  Table~\ref{tab:linelist} presents the entire line list.


With the final line list, the mean absolute deviation between the
pixels of the solar spectrum and the pixels of its synthesis is $5.6
\times 10^{-3}$, and the standard deviation is $1.3 \times 10^{-2}$.
For \arcturus, the mean absolute deviation is $9.4 \times 10^{-3}$,
and the standard deviation is $2.1 \times 10^{-2}$.  The units are
such that the continuum is 1.

\subsection{Synthetic spectra generation}
\label{sec:specgen}
We generate the library of synthetic spectra from the final line list.
Each spectrum ranges from 6300~\AA\ to 9100~\AA\ with a resolution of
0.02~\AA.  To save computation time in the abundance determination, we
bin each synthetic spectrum by a factor of 7 (0.14~\AA\ resolution).
When the binned and unbinned spectra are smoothed to the best
resolution provided by \deimos\ ($\sim 1.1$~\AA\ FWHM), individual
pixels in the binned spectra differ from the unbinned spectra by less
than one part in $10^3$.

\subsection{\feh\ and \afe\ determination}
\label{sec:abundance}
Our continuum normalization for observed spectra (see
\S\,\ref{sec:continuum}) may be depressed very slightly by weak
absorption lines.  To counteract this effect, we allow weak absorption
in synthetic spectra to affect their normalizations in the same way
that they affect observed spectra.  First, we interpolate a synthetic
spectrum onto the same wavelength array as the observed spectrum.
Then, we smooth the synthetic spectrum through a Gaussian filter whose
width is the observed spectrum's measured spectral resolution as a
function of wavelength (see \S\,\ref{sec:reduction}).  To complete the
renormalization, we divide the synthetic spectrum by the synthetic
``continuum'' determined exactly the same way as in
\S\,\ref{sec:continuum}, including the continuum regions and weighting
by the inverse variance of the \emph{observed} spectrum.

We discard all stars with photometric $\mathteff < 4000$~K because
such stars are susceptible to TiO absorption, which we do not attempt
to model.  The coolest temperature on the synthetic grid is 4000~K for
this reason.

Next, we compute $\chi^2$ for an initial guess at the four atmospheric
parameters \teff, \logg, \feh, and \afe.  A Levenberg-Marquardt
optimization algorithm finds the values within the bounds of the grid
that minimize $\chi^2$ computed from the difference between the
observed spectrum and trial synthetic spectrum.  We sample the
parameter space between grid points by linearly interpolating the
synthetic spectra at the neighboring grid points.  The parameters for
the best-fit spectrum are the final atmospheric parameters for the
star.  The fitting error on each parameter is $\sqrt{\sigma^2 \chi^2}$
where $\sigma^2$ is the parameter's diagonal element of the covariance
matrix.  The fitting errors are usually much smaller than the total
error, including systematic error, which will be discussed in
\S\,\ref{sec:error}.  Fitting errors are particularly small for high
S/N spectra.

Finding the global minimum $\chi^2$ in four dimensions is difficult.
Surface gravity is particularly difficult to measure.  First, the red
spectral region contains very few gravity sensitive lines that are
easy to model.  Second, few transitions from ionized species are
visible in our spectra, making surface gravity poorly constrained.  To
reduce the dimensionality of the parameter space, we fix \teff\ and
\logg\ based on photometry (P. B. Stetson, private communication) and
theoretical isochrones shifted to the distance modulus of the target.
MRS abundances almost always agree with HRS abundances more closely
when we fix \teff\ and \logg\ photometrically.  \citet[][revised
2003]{har96} provides the distance modulus and reddening of each
cluster.

For the GCs in this article, we use 14.0~Gyr, $\mathafe = +0.3$
isochrones from the Yonsei-Yale group \citep[YY,][]{dem04}.  \teff\
and \logg\ are mostly insensitive to age and $\alpha$ enhancement.
Given a star's $I$ magnitude and $V-I$ color, we linearly interpolate
between tracks of constant age to recover \teff\ and \logg.  (We do
not have $I$ measurements for a few stars in \ngcc\ nor any star in
\ngcg.  Instead, we use $V$ and $B-V$.)  We have also experimented
with Victoria-Regina \citep{van06} and Padova \citep{gir02}
isochrones.  Although the Padova photometric \feh\ is quite different
from the other two sets of isochrones near the tip of the RGB, the
photometric \teff\ and \logg\ do not change enough to affect abundance
analysis.  We have also compared theoretical YY temperatures to the
empirical temperatures of \citet[][RM05]{ram05}.  For $\mathteff \la
4800$~K, the average discrepancy $\Delta\mathteff < 50$~K.  For
$\mathteff \ga 5000$~K, the RM05 \teff\ tends to be up to 250~K lower
than the YY \teff.  Given their success in reproducing GC abundances
(\S\,\ref{sec:feh}), we have chosen to use YY temperatures.

Future applications of this method to inhomogeneous stellar
populations may require an assumption of the age or iteration of the
age until the spectroscopic \feh\ agrees with the photometric \feh.
In theory, this iteration will give the age of the star.  However,
photometric errors and the systematic errors in both abundance
measurement methods will undoubtedly make such an age very uncertain.
In reality, \teff\ and \logg\ change little with age or metallicity
for a given $V-I$ color, so the assumed isochrone parameters hardly
affect the abundance results.


\section{Results and comparison to HRS abundances}
\label{sec:results}
To demonstrate the effectiveness of the MRS abundance technique, we
have observed GCs specifically to compare MRS to HRS metallicities and
$\alpha$ enhancements.  In this section, we present those comparisons
and examine MRS results for systematic trends.

\subsection{\teff}
\begin{figure}
\plotone{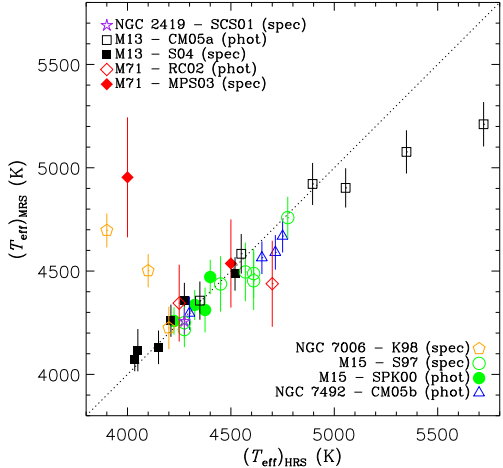}
\caption{\teff\ determined from MRS without photometric information
  for individual stars (this work) vs.\ \teff\ used in HRS studies of
  the same stars.  $(\mathteff)_{\mathrm{HRS}}$ is not necessarily
  determined spectroscopically.  The figure legend indicates whether
  $(\mathteff)_{\mathrm{HRS}}$ is determined spectroscopically
  (``spec'') or photometrically (``phot'').  At $\mathteff > 5000$~K,
  MRS analysis may underpredict \teff\ because at high temperatures,
  lines with low EPs become immeasurably weak.  Alternatively, the
  photometric temperatures for M13 based on theoretical atmospheres
  \citep{coh05a} may be overestimates.  The dotted line is one-to-one.
  The following references provide $(\mathteff)_{\mathrm{HRS}}$:
  SCS01: \citet{she01}, CM05a: \citet{coh05a}, S04: \citet{sne04},
  RC02: \citet{ram02}, MPS03: \citet{mis03}, K98: \citet{kra98}, S97:
  \citet{sne97}, SPK00: \citet{sne00}, and CM05b:
  \citet{coh05b}.\label{fig:teff}}
\end{figure}

The MRS technique can recover \teff\ even without photometry.
Figure~\ref{fig:teff} shows the \teff\ of the best-fit atmospheric
parameters when we allow \teff, \logg, \feh, and \afe\ to vary
compared to the published \teff\ used in HRS studies of the same
stars.  Not all authors of HRS studies determine \teff\
spectroscopically.  The figure legend indicates which HRS studies
choose \teff\ such that lines of different excitation potential yield
the same abundance (``spec'') and which HRS studies rely exclusively
on broadband photometry to determine \teff\ (``phot'').

$(\mathteff)_{\mathrm{MRS}}$ reproduces $(\mathteff)_{\mathrm{HRS}}$
very well in most cases.  The absolute deviation for 77\% of stars
falls within 150~K.  At $\mathteff > 5000$~K, MRS analysis
underpredicts \teff\ because at high temperatures, lines with low EPs
become immeasurably weak.  [An alternative explanation is that the
photometric temperatures of \citet{coh05a} could be inaccurate.]
Spectroscopic temperature is measured by comparing the strengths of
lines with a range of EP.  Without the low-EP lines, the temperature
is difficult to measure.  Fixing \teff\ photometrically alleviates
this problem and also eliminates the large random error in a
spectroscopic \teff\ of lower S/N stars.  For the remainder of this
article, both \teff\ and \logg\ are set by photometry.

\subsection{\feh}
\label{sec:feh}
\begin{figure}
\plotone{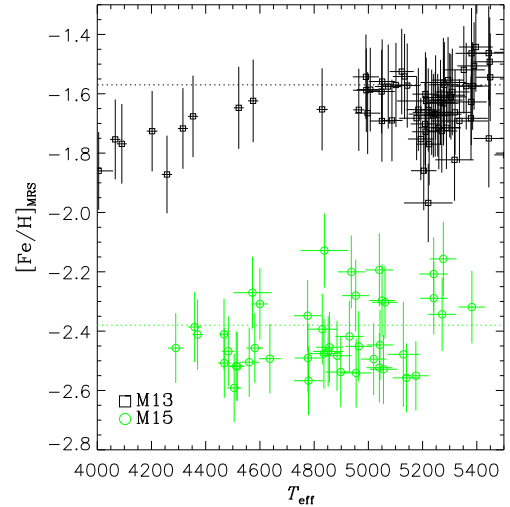}
\caption{$\mathfeh_{\mathrm{MRS}}$ vs.\ \teff\ determined from $V-I$
  color of individual stars in M13 and M15.  (A few stars in M13 rely
  on $B-V$ rather than $V-I$.)  The lack of a trend suggests little
  systematic covariance between measurements of
  $\mathfeh_{\mathrm{MRS}}$ and \teff.  The dotted lines show
  $\mathfeh_{\mathrm{HRS}}$ \citep{pri05}.\label{fig:teffah}}
\end{figure}

We expect any spread in \feh\ within a single GC to be random error
because these GCs are monometallic.  Correlation of \feh\ with other
parameters can indicate systematic errors.  First, we show in
Fig.~\ref{fig:teffah} the spectroscopic \feh\ vs.\ photometric
\teff\ in \ngcc\ and \ngcf.  In all of \ngcf\ and above 5000~K in
\ngcc, we see only random scatter.  The bright, cool stars near the
tip of the RGB in \ngcc\ show a slight positive slope of $\sim
0.1$~dex from 4000~K to 5000~K.  The covariance is undesirable but not
unexpected.  Both higher temperatures and lower metallicities weaken
absorption lines.  The magnitude and significance of the trend are
small and restricted to the hottest stars.  Furthermore, we emphasize
that \ngcc\ is the \emph{worst} case.  No other GC shows this
correlation.

\begin{figure}
\plotone{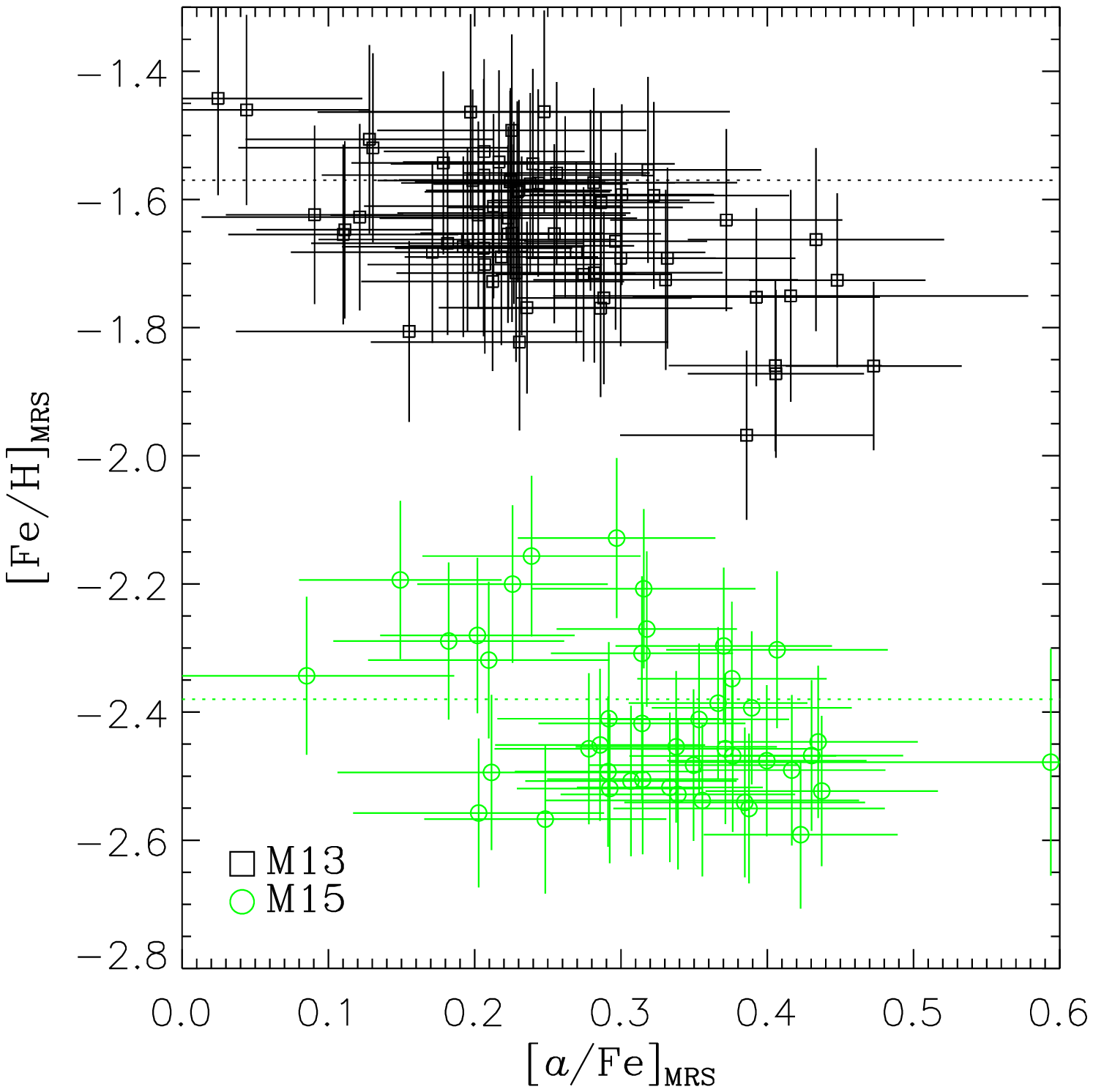}
\caption{$\mathfeh_{\mathrm{MRS}}$ vs.\ $\mathafe_{\mathrm{MRS}}$ of
  individual stars in M13 and M15.  The lack of a trend suggests
  little systematic covariance between measurements of
  $\mathafe_{\mathrm{MRS}}$ vs.\ $\mathfeh_{\mathrm{MRS}}$.  The
  dotted lines show $\mathfeh_{\mathrm{HRS}}$
  \citep{pri05}.\label{fig:alphaah}}
\end{figure}

We also expect no correlation between \afe\ and \feh\ within a single
GC.  Both \afe\ and \feh\ affect the strength of $\alpha$ element
absorption lines.  If there were a systematic trend, we would expect
the two parameters to be anti-correlated.  No cluster shows a
convincing anti-correlation.  Figure~\ref{fig:alphaah} shows both
parameters for \ngcc\ and \ngcf.  \ngcc\ is again the worst case, and
the significance of any trend is destroyed by removing the one or two
points with the most discrepant \feh.

\begin{figure}
\plotone{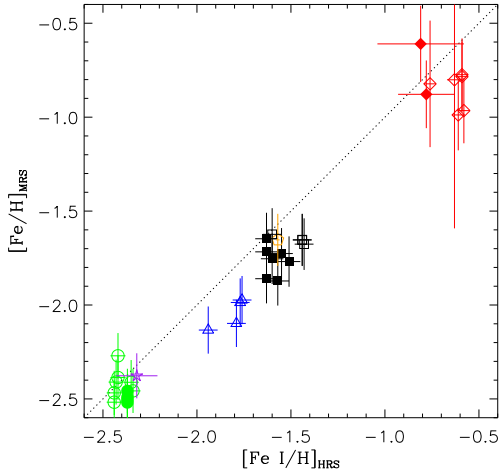}
\caption{This work's \feh\ vs.\ [\ion{Fe}{1}/H] determined from
  neutral iron lines in HRS.  The dotted line is one-to-one.  The
  symbols, colors, and references are the same as in
  Fig.~\ref{fig:teff}.\label{fig:fei}}
\end{figure}

The MRS technique matches the metallicities determined from HRS to
within the HRS metallicity scatter for a given GC.
Figure~\ref{fig:fei} shows \feh\ of the best-fit atmospheric
parameters (where photometry determines \teff\ and \logg) compared to
$\mathfeh_{\mathrm{HRS}}$ determined from \ion{Fe}{1} lines in the
range $-2.38 < \mathfeh < -0.76$.  We choose \ion{Fe}{1} because
\ion{Fe}{1} lines greatly outnumber \ion{Fe}{2} lines in the \deimos\
spectra with our range of \teff\ and \logg.  Although these GCs are
almost certainly monometallic, we have plotted the measurements for
individual stars on both axes.  The range of $\mathfeh_{\mathrm{HRS}}$
within a single GC, particularly depending on the authors of the
measurement, demonstrates the uncertainty to which the mean
metallicity is known.  The MRS measurements do not fall outside of
this scatter.

\begin{figure}
\plotone{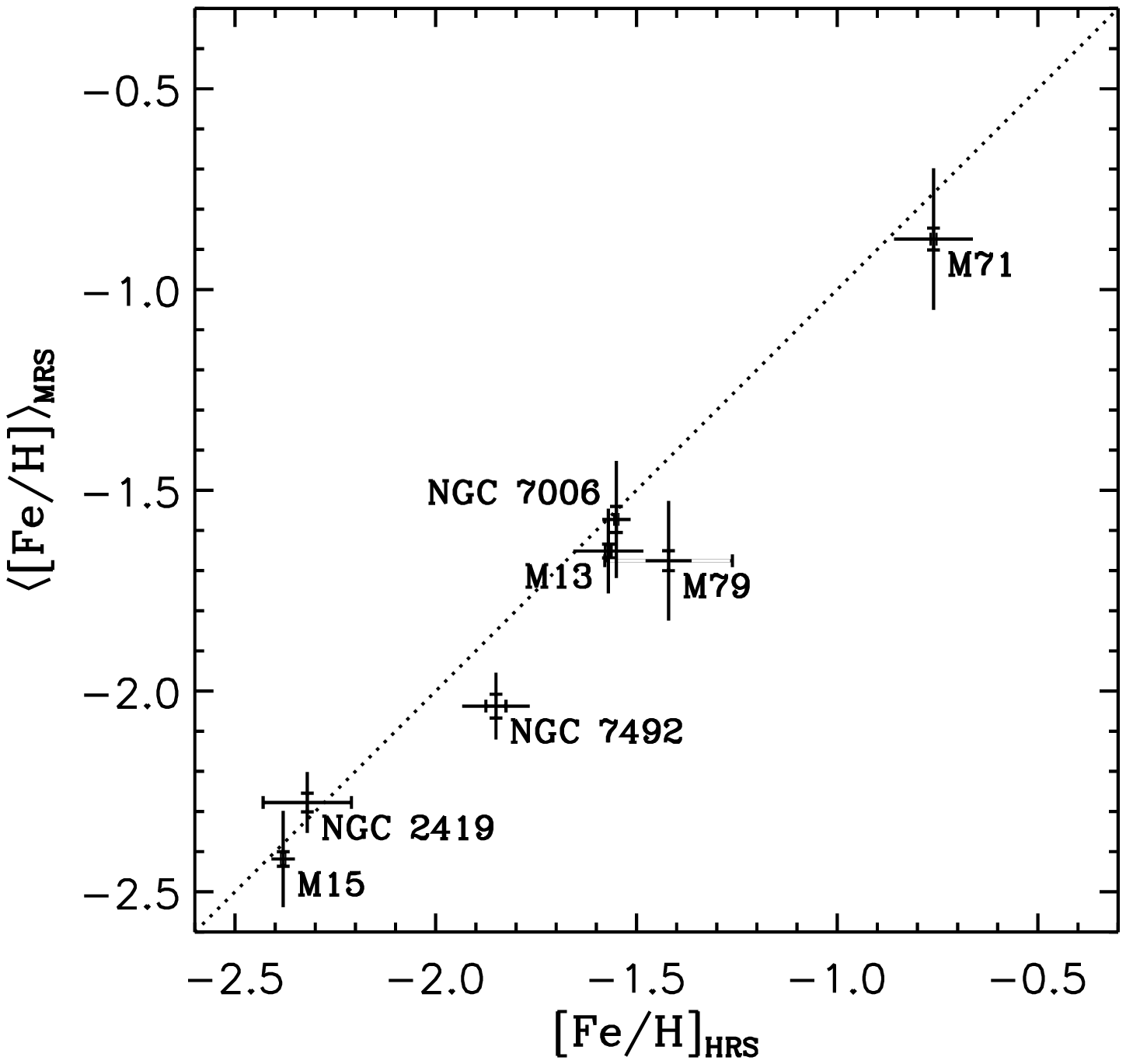}
\caption{Mean \feh\ values in GCs observed with \deimos\ vs.\ values
  of \feh\ determined from HRS \citep{pri05}.  The dotted line is
  one-to-one.  The full error bars are weighted sample standard
  deviations, and the hash marks along the error bars are weighted
  errors on the mean.  The HRS errors are weighted by the inverse
  square of the measurement errors of all of the individual stars in
  the following references.  M79: \citet{gra89}, NGC~2419 (the HRS
  error bar is the individual measurement error for the only star
  available): \citet{she01}, M13: \citet{sne04} and \citet{coh05a},
  M71: \citet{gra86}, \citet{sne94}, \citet{ram02}, and
  \citet{mis03}, NGC 7006: \citet{kra98}, M15: \citet{sne97} and
  \citet{sne00}, NGC~7492: \citet{coh05b}\label{fig:fecompare}}
\end{figure}

\begin{deluxetable*}{lcccccccc}
\tablewidth{0pt}
\tablecolumns{9}
\tablecaption{Average Globular Cluster Abundances Comparison\label{tab:results}}
\tablehead{\colhead{Cluster}  &  \colhead{$N_{\mathrm{HRS}}$\tablenotemark{a}}  &  \colhead{$\mathfeh_{\mathrm{PVI05}}$\tablenotemark{b}}  &  \colhead{$\langle\mathafe_{\mathrm{HRS}}\rangle$\tablenotemark{c}}  &  \colhead{$N_{\mathrm{DEIMOS}}$\tablenotemark{d}}  &  \colhead{$\langle\mathfeh_{\mathrm{MRS}}\rangle$}  &  \colhead{$\mathrm{rms}\,(\mathfeh_{\mathrm{MRS}})$}  &  \colhead{$\langle\mathafe_{\mathrm{MRS}}\rangle$}  &  \colhead{$\mathrm{rms}\,(\mathafe_{\mathrm{MRS}})$}}
\startdata
\ngca  & \phn2 & $-1.42$ & $+0.26 \pm 0.08$ & 33 & $-1.69 \pm 0.02$ & 0.15 & $+0.27 \pm 0.01$ & 0.11 \\
\ngcb  & \phn1 & $-2.32$ & $+0.20 \pm 0.08$ & 30 & $-2.28 \pm 0.02$ & 0.08 & $+0.28 \pm 0.01$ & 0.09 \\
\ngcc  &    60 & $-1.57$ & $+0.20 \pm 0.01$ & 69 & $-1.66 \pm 0.02$ & 0.11 & $+0.25 \pm 0.01$ & 0.09 \\
\ngcd  &    40 & $-0.76$ & $+0.27 \pm 0.01$ & 47 & $-0.92 \pm 0.04$ & 0.17 & $+0.27 \pm 0.01$ & 0.17 \\
\ngce  & \phn6 & $-1.55$ & $+0.24 \pm 0.01$ & 20 & $-1.59 \pm 0.03$ & 0.13 & $+0.35 \pm 0.01$ & 0.08 \\
\ngcf  &    49 & $-2.38$ & $+0.42 \pm 0.03$ \tablenotemark{e}& 44 & $-2.42 \pm 0.01$ & 0.12 & $+0.33 \pm 0.01$ & 0.07 \\
\ngcg  & \phn4 & $-1.85$ & $+0.23 \pm 0.02$ & 21 & $-2.04 \pm 0.02$ & 0.08 & $+0.32 \pm 0.02$ & 0.10 \\
\enddata
\tablecomments{The rms values and errors on the mean are weighted by
  the inverse square of individual measurement errors.}
\tablenotetext{a}{Number of stars observed with high resolution
  spectroscopy.}
\tablenotetext{b}{\citet{pri05}, high resolution spectroscopy.}
\tablenotetext{c}{See Fig.~\ref{fig:fecompare} for references from
  which these averages were calculated.}
\tablenotetext{d}{Number of \deimos\ spectra analyzed in this
  article.}
\tablenotetext{e}{See \S\,\ref{sec:afe} for a discussion of the
  anomalously large value of $\mathafe_{\mathrm{HRS}}$ for \ngcf.}
\end{deluxetable*}

We measure individual metallicities for all stars that we observe.  We
eliminate non-members by radial velocity, and we discard stars with
photometric \teff\ outside the range of the spectral grid.  We also
discard HB stars and stars with noticeable TiO absorption.
Figure~\ref{fig:fecompare} shows the weighted mean of
$\mathfeh_{\mathrm{MRS}}$ of all the stars in each cluster.  The
abscissa is $\mathfeh_{\mathrm{HRS}}$ from the compilation of
\citet{pri05}.  Two types of errors bars are shown on both axes: the
standard error on the mean and the sample standard deviation.  The
latter represents the typical measurement error on one star.  The HRS
error bars are determined from $\mathfeh_{\mathrm{HRS}}$ of individual
stars in the references given in the figure caption.  The error bars
for both MRS and HRS are weighted by the inverse square of the
individual measurement errors.  Table~\ref{tab:results} lists the same
data along with the number of stars in each sample.  The HRS
measurements for \ngcb, \ngca, and \ngcg\ are based respectively on
only 1, 2, and 4 stars.  We suggest that, in these cases, a cluster's
mean \feh\ may be more precisely determined with our larger MRS
samples.

Table~\ref{tab:mrsdata} lists MRS results for individual stars.  Where
available, all three photometric magnitudes ($B$, $V$, and $I$) are
given, but only two determine \teff\ and \logg.  $V$ and $I$ are
preferred.  \teff\ and \logg\ in this table are the values that we
have determined photometrically.  The last three columns are data from
and references to HRS studies for the stars in common between the MRS
and HRS datasets.

All GCs seem to display internal variations in [C/H] and [N/H]
\citep[e.g.,][]{gru99}.  \citet{coh05} show ranges of 2--3 dex of
[N/H] in five GCs, including \ngcc, \ngcd, and \ngcf.  The variations
could alter the strength of CN absorption.  \ngcd\ is the only GC in
our sample metal-rich enough to exhibit strong red CN absorption.  If
star-to-star CN abundance variations affect $\mathfeh_{\mathrm{MRS}}$,
then the metallicity determined only from the spectral region affected
by CN (7850--8400~\AA) should differ from the metallicity determined
from the rest of the spectrum.  We subjected \ngcd\ to this test, and
\feh\ measurements from the two cases agree very closely for every
star.  CN abundance variations, if they exist, do not appear to
contribute to the error on individual stars or to the large scatter in
$\mathfeh_{\mathrm{MRS}}$ for \ngcd.

\subsection{\afe}
\label{sec:afe}
\begin{figure}
\plotone{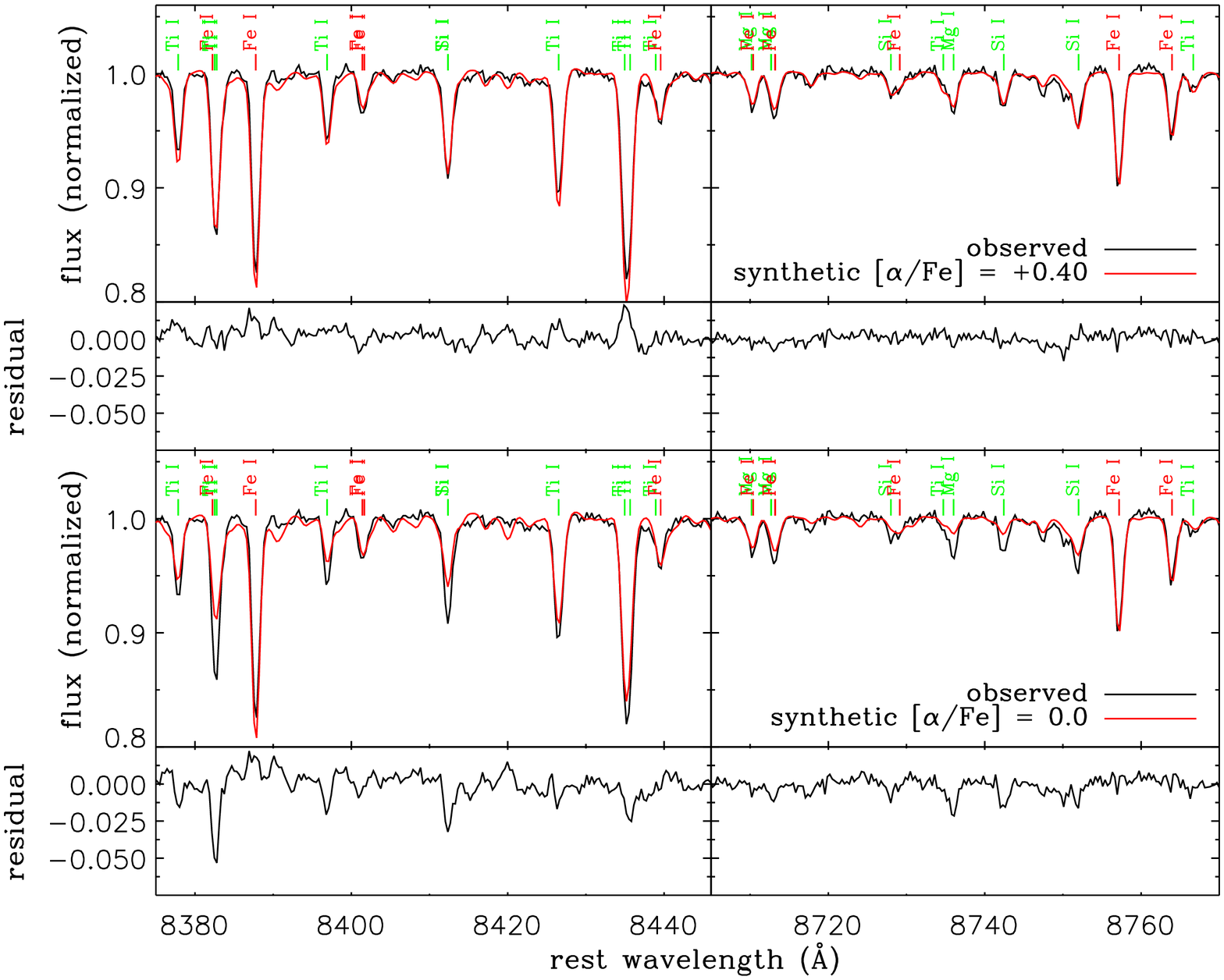}
\caption{{\it Top:} Portions of an observed spectrum of a star in M13
  (black) and the best fit synthetic spectrum (red), which has
  $\mathafe = +0.40$.  Fe lines are labeled in red and $\alpha$
  element lines are labeled in green.  The residual is the difference
  between the observed and synthetic spectra.  {\it Bottom:} The same
  observed spectrum (black) and a synthetic spectrum (red) with the
  same parameters as the synthetic spectrum in the top panel except
  that $\mathafe = 0.0$.  The observed spectrum is highly inconsistent
  with the solar value of \afe.\label{fig:alpha_solar}}
\end{figure}

MRS can give an abundance dimension beyond metallicity.  Most of the
stellar absorption lines visible in \deimos\ spectra are from
\ion{Fe}{1}, but there are a comparable number of $\alpha$ element
absorption lines.  By far, the most abundant are Ti, Si, Ca, and Mg,
in order of decreasing prevalence.  Figure~\ref{fig:alpha_solar} shows
two spectral regions with high concentrations of lines from all of
these elements except Ca.  The figure also shows two synthetic spectra
that are identical except for their values of \afe.  The observed
spectrum is consistent with the synthetic spectrum for which $\mathafe
= +0.40$, but highly inconsistent with the synthetic spectrum for
which $\mathafe = 0.0$.  The discrepancy demonstrates that MRS can
easily distinguish between the halo plateau and solar values of \afe.

\begin{figure*}
\plotone{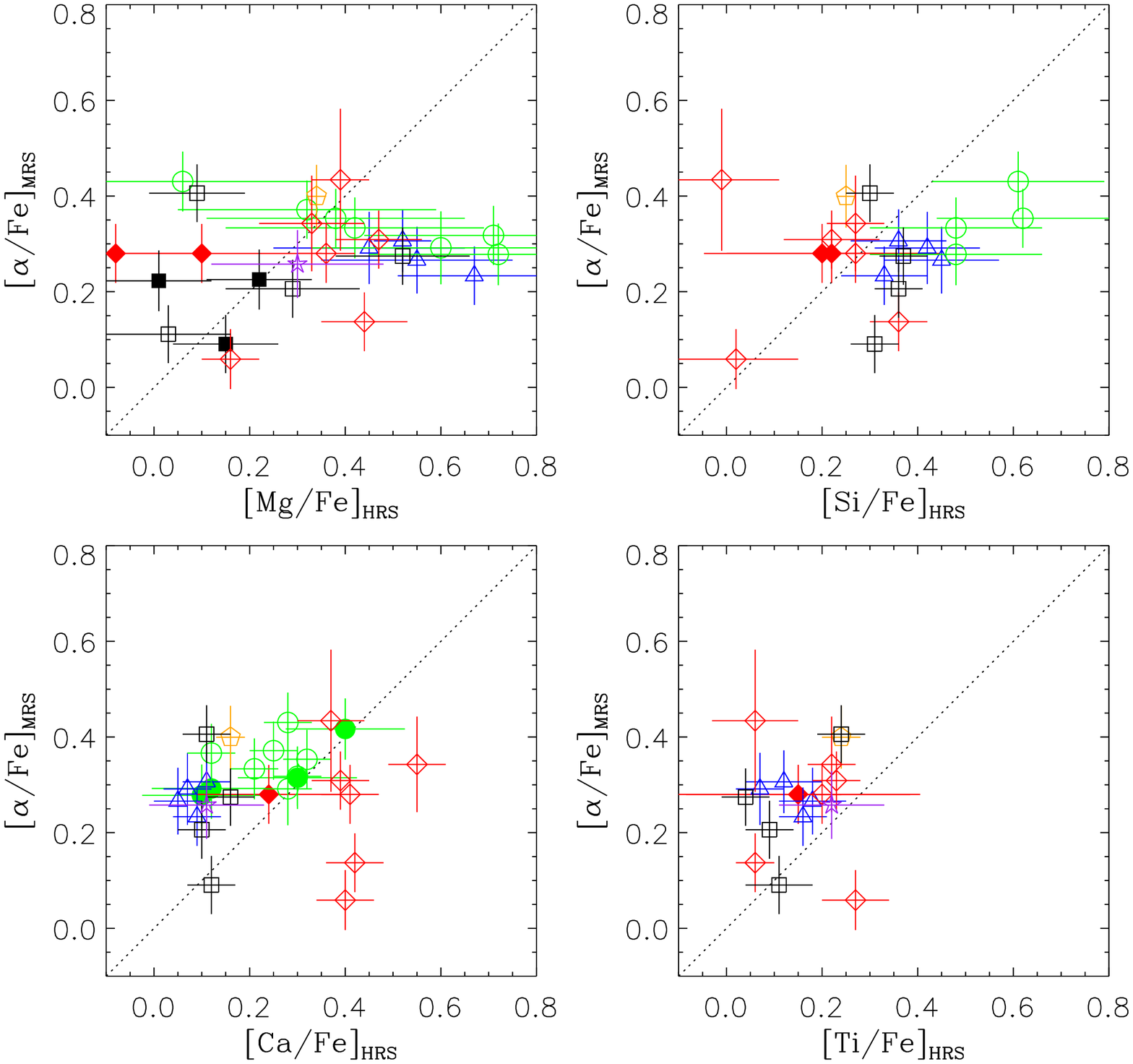}
\caption{$\mathafe_{\mathrm{MRS}}$ vs.\ individual $\alpha$ element
  abundances determined from HRS.  The dotted lines are one-to-one.
  The symbols, colors, and references are the same as in
  Fig.~\ref{fig:teff}.\label{fig:alpha}}
\end{figure*}

In order to build additional confidence in the ability of MRS to
determine \afe, we compare values of $\mathafe_{\mathrm{MRS}}$ to HRS
measurements.  Figure~\ref{fig:alpha} shows $\mathafe_{\mathrm{MRS}}$
compared to HRS measurements of [Mg/Fe], [Si/Fe], [Ca/Fe], and
[Ti/Fe].  All four plots show weak or no correlations.  The lackluster
agreement is not surprising because $\mathafe_{\mathrm{MRS}}$ is a
weighted combination of all four elements.

To better compare MRS and HRS results, we define
$\mathafe_{\mathrm{HRS}}$, which is a weighted mean of the four
$\alpha$ element ratios.

\begin{figure}
\plotone{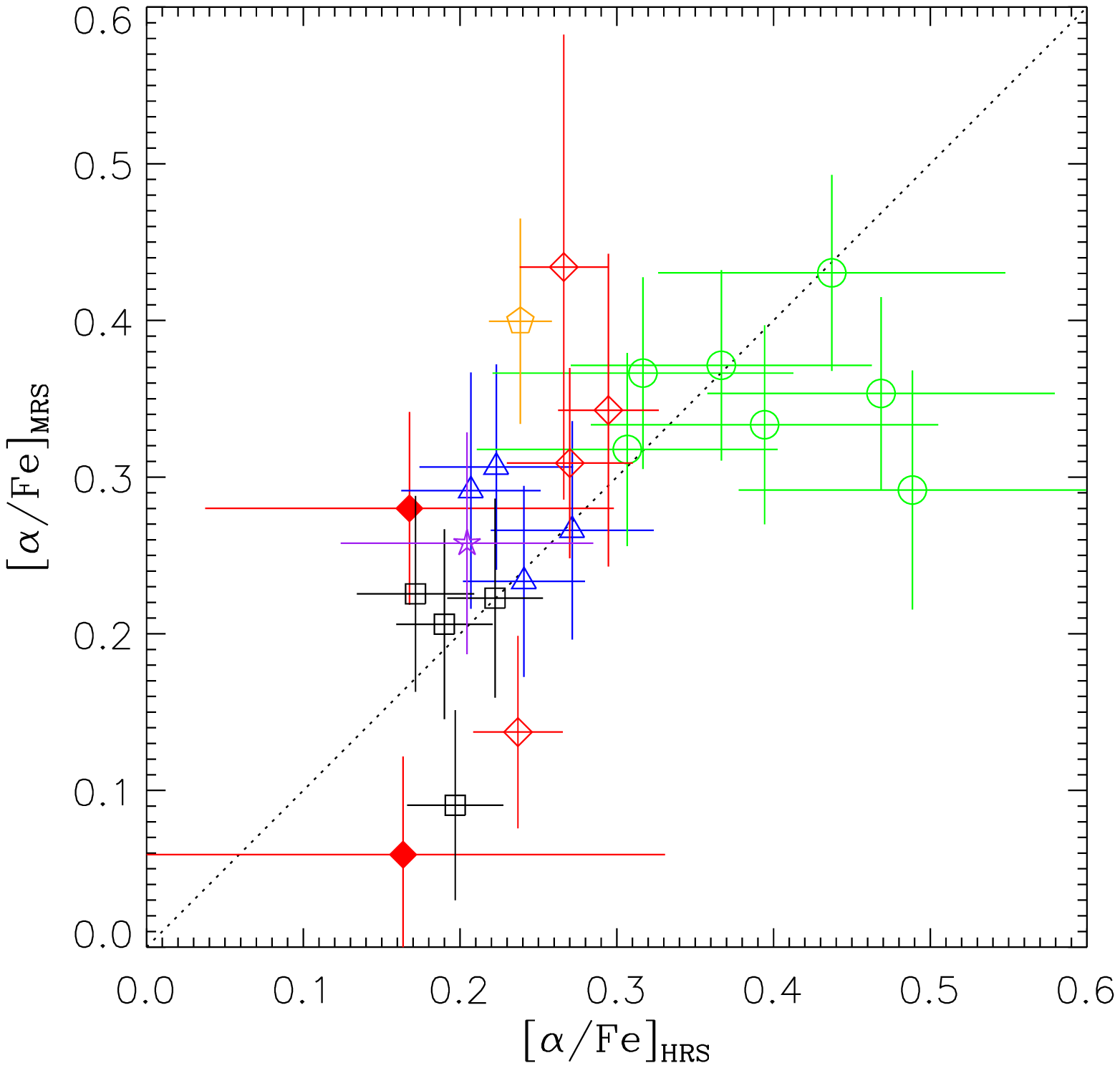}
\caption{$\mathafe_{\mathrm{MRS}}$ vs.\ a weighted average of the
  available [Mg/Fe], [Si/Fe], [Ca/Fe], and [Ti/Fe] HRS measurements
  for individual stars.  Eqs.~\ref{eq:afe_hrs} and \ref{eq:afeerr_hrs}
  give the formulas for deriving $\mathafe_{\mathrm{HRS}}$ and its
  error from the individual element measurements.  The dotted line is
  one-to-one.  The symbols, colors, and references are the same as in
  Fig.~\ref{fig:teff}.  See \S\,\ref{sec:afe} for a discussion of the
  large values of $\mathafe_{\mathrm{HRS}}$ for M15 (green
  points).\label{fig:alpha_avg}}
\end{figure}

\begin{eqnarray}
\label{eq:afe_hrs}
\mathafe_{\mathrm{HRS}} &=& (\gamma_{\mathrm{Mg}}\,\mathrm{[Mg/Fe]}+4\gamma_{\mathrm{Si}}\,\mathrm{[Si/Fe]}\\
 & & +2\gamma_{\mathrm{Ca}}\,\mathrm{[Ca/Fe]}+6\gamma_{\mathrm{Ti}}\,\mathrm{[Ti/Fe]}) \nonumber\\
 & & \times (\gamma_{\mathrm{Mg}}+4\gamma_{\mathrm{Si}}+2\gamma_{\mathrm{Ca}}+6\gamma_{\mathrm{Ti}})^{-1}\nonumber\\
\label{eq:afeerr_hrs}
\delta\mathafe_{\mathrm{HRS}} &=& \big[\gamma_{\mathrm{Mg}}\left(\delta\mathrm{[Mg/Fe]}\right)^2+\gamma_{\mathrm{Si}}\left(4\,\delta\mathrm{[Si/Fe]}\right)^2\\
 & & +\gamma_{\mathrm{Ca}}\left(2\,\delta\mathrm{[Ca/Fe]}\right)^2+\gamma_{\mathrm{Ti}}\left(6\,\delta\mathrm{[Ti/Fe]}\right)^2\big]^{1/2} \nonumber\\
 & & \times (\gamma_{\mathrm{Mg}}+4\gamma_{\mathrm{Si}}+2\gamma_{\mathrm{Ca}}+6\gamma_{\mathrm{Ti}})^{-1}\nonumber
\end{eqnarray}

\noindent
The $\gamma_{\mathrm{X}}$ coefficients are 1 if the element X has a
published HRS value and 0 otherwise.  $\delta\mathrm{[X/Fe]}$ is the
published error.  For those studies that quote [X/H] instead of
[X/Fe], $\delta\mathrm{[X/Fe]} = \sqrt{(\delta\mathrm{[X/H]})^2 +
(\delta\mathrm{[Fe/H]})^2}$.  The weights 1, 4, 2, and 6 mimic the
relative presence of absorption from the four elements.  This
combination of weights also gives good correlation between
$\mathafe_{\mathrm{MRS}}$ and $\mathafe_{\mathrm{HRS}}$, shown in
Fig.~\ref{fig:alpha_avg}.  These data are also listed in
Table~\ref{tab:mrsdata}.  Although the weights in
$\mathafe_{\mathrm{HRS}}$ approximate the weights that the individual
elements receive in determining $\mathafe_{\mathrm{MRS}}$, matching
the elemental averages exactly is not possible.  Therefore, the
comparisons between $\mathafe_{\mathrm{MRS}}$ and
$\mathafe_{\mathrm{HRS}}$ are approximate checks of agreement.

Three studies deserve particular mention.  The \ngcf\ measurements of
\citet{sne00} use Hydra spectra, which have lower spectral resolution
than most HRS studies.  They do not attempt to measure Mg, and most of
their stars do not display enough Si or Ti absorption to permit
accurate measurements of those elemental abundances.  Furthermore,
their Ca measurements are based on only one line.  The \ngcc\
measurements of \citet{sne04} do not include Si, Ca, or Ti.  They do
measure Mg, but Mg is the least visible $\alpha$ element in the
\deimos\ spectra.  For these reasons, we exclude the
$\mathafe_{\mathrm{HRS}}$ measurements of \citet{sne00} and
\citet{sne04} from Fig.~\ref{fig:alpha_avg} and
Table~\ref{tab:mrsdata}.  We also draw attention to the \ngcf\
measurements of \citet{sne97}.  They measure particularly large values
for [Si/Fe], and they do not recover [Ti/Fe] for any of the stars that
are in common between their HRS sample and our MRS sample.  Therefore,
$\mathafe_{\mathrm{HRS}}$ has a different meaning for the \ngcf\ data
(green points in Fig.~\ref{fig:alpha_avg}) than for the data from
other clusters.

\begin{figure}
\plotone{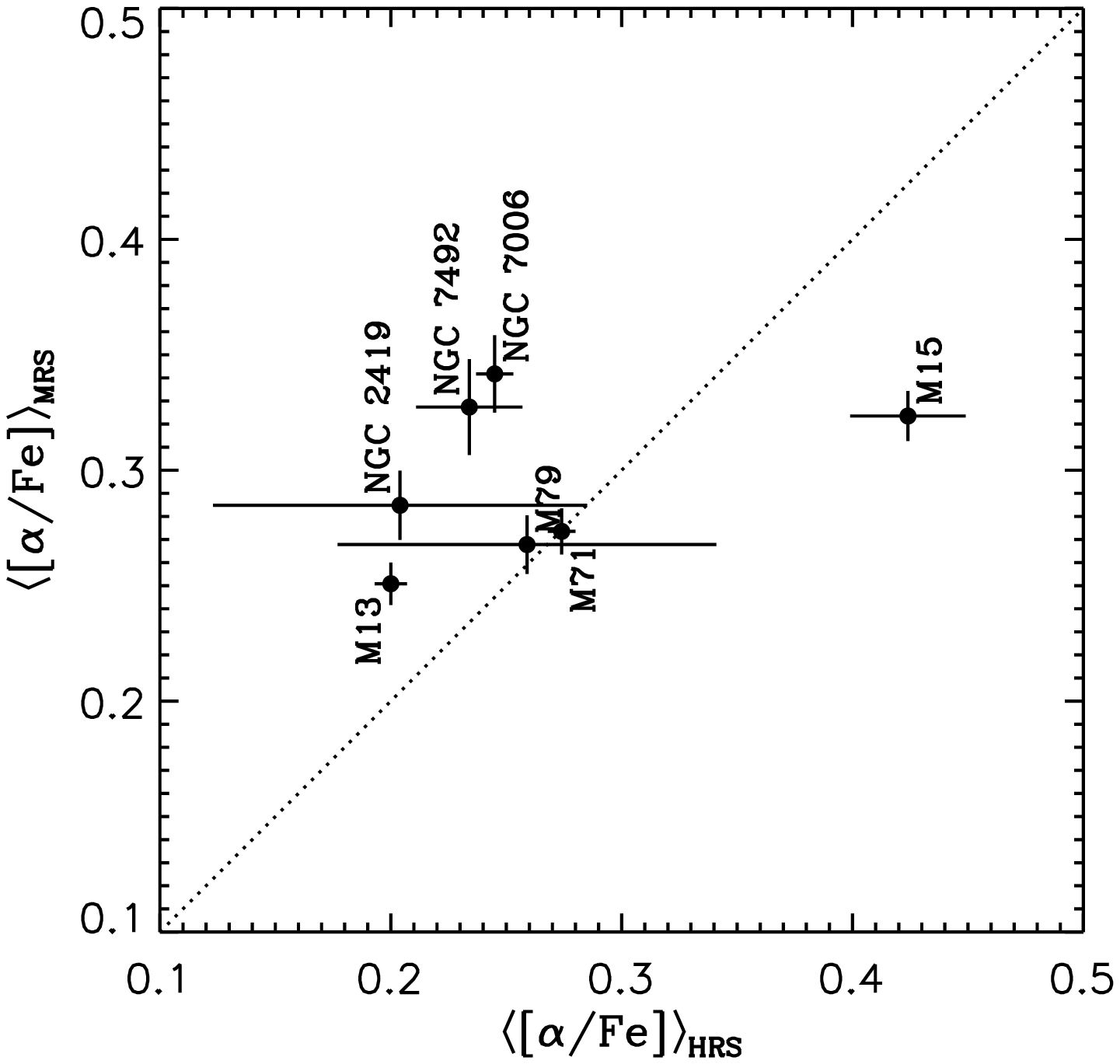}
\caption{Mean cluster values of $\mathafe_{\mathrm{MRS}}$ vs.\
  $\mathafe_{\mathrm{HRS}}$.  The range of \afe\ in these GCs is very
  small, but the points are roughly consistent with the one-to-one
  line (dotted line).  See \S\,\ref{sec:afe} for a discussion of the
  large value of $\langle\mathafe_{\mathrm{HRS}}\rangle$ for
  M15.\label{fig:alpha_alpha}}
\end{figure}

Comparing individual stars limits the sample to stars observed with
both MRS and HRS.  In order to draw on the full set of stars observed
with \emph{either} MRS or HRS, we compute mean values of \afe\ for
entire clusters.  $\langle\mathafe\rangle$ is weighted by the inverse
square of $\delta\mathafe$.  Figure~\ref{fig:alpha_alpha} shows the
comparison.  These averages are also listed in
Table~\ref{tab:results}.  The range of $\langle\mathafe\rangle$ for
these GCs is very small, but $\langle\mathafe\rangle_{\mathrm{MRS}}$
and $\langle\mathafe\rangle_{\mathrm{HRS}}$ agree to $\la 0.1$~dex for
all clusters.

\begin{figure}
\plotone{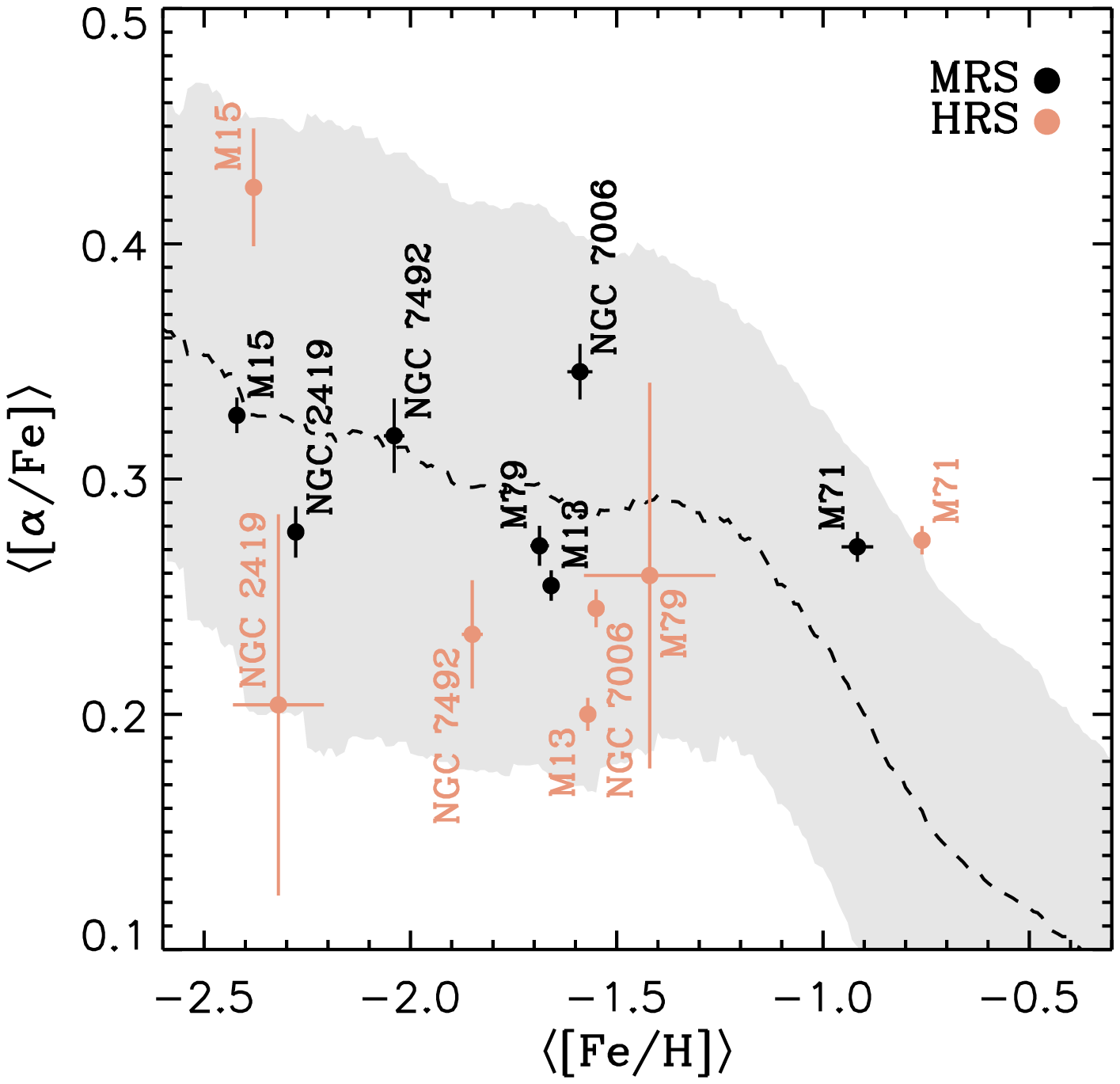}
\caption{Mean cluster values of \afe\ vs.\ \feh\ for MRS (black) and
  HRS (pink).  For comparison, we show the relation for MW stars from
  \citet{ven04}.  The dashed line is the average \afe\ in a moving
  window with a half width of 0.5~dex in \feh, and the shaded region
  is the rms spread.  $\mathafe_{\mathrm{MRS}}$ is very near the MW
  halo plateau value of $\mathafe = +0.3$ for every
  cluster.\label{fig:alphafe_feh}}
\end{figure}

Finally, we show $\langle\mathafe\rangle$ vs.\
$\langle\mathfeh\rangle$ in Fig.~\ref{fig:alphafe_feh}.  Also shown is
the relationship between these two quantities for MW stars tabulated
by \citet{ven04}.  The $\langle\mathfeh\rangle$ values from both MRS
and HRS are consistent with the halo plateau value of $\mathafe \sim
+0.3$.


\section{Quantification of Abundance Errors}
\label{sec:error}
\subsection{Total Error on \feh}
\label{sec:fehtoterr}
The Levenberg-Marquardt algorithm which determines the best-fit
synthetic spectrum by minimizing $\chi^2$ gives an estimate of the
fitting error based on the depth of the $\chi^2$ minimum in parameter
space.  However, the fitting error is usually a small part of the
total error.  The major sources of error in high S/N spectra are
errors in atmospheric parameters and imperfect spectral modeling.  Two
effects make the total error a function of \feh.  First, higher
metallicity spectra are more sensitive to errors in \teff\ and \logg\
(see \S\,\ref{sec:atmerr}).  Second, higher metallicity spectra
exhibit more complex absorption from molecular transitions not seen at
lower metallicity.

In order to quantify the total error on $\mathfeh_{\mathrm{MRS}}$, for
each cluster we find the systematic error
$\delta_{\mathrm{sys}}\mathfeh_{\mathrm{MRS}}$ that satisfies

\begin{figure}
\plotone{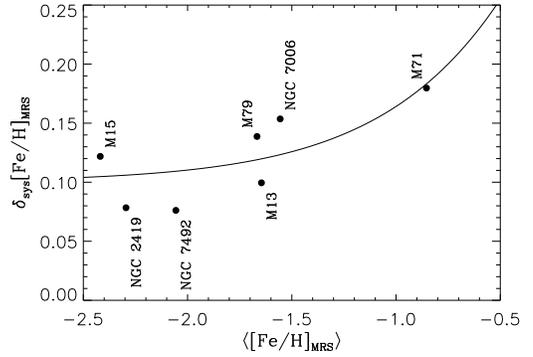}
\caption{The systematic error in \feh\ as a function of the mean \feh\
  for each cluster.  The line is the best fit of the function $\ln
  \left(e^{0.1} + e^{a + bx}\right)$.\label{fig:feherr}}
\end{figure}

\begin{equation}
\mathrm{rms}\,\left(\frac{\mathfeh_{\mathrm{MRS}} -
  \langle\mathfeh_{\mathrm{MRS}}\rangle}{\sqrt{\left(\delta_{\mathrm{fit}}\mathfeh_{\mathrm{MRS}}\right)^2
    + \left(\delta_{\mathrm{sys}}\mathfeh_{\mathrm{MRS}}\right)^2}}\right) = 1\:.
\end{equation}

\noindent
Figure~\ref{fig:feherr} shows the results.
$\delta_{\mathrm{sys}}\mathfeh_{\mathrm{MRS}}$ appears to approach 0.1
at low \feh\ and increase as \feh\ rises above $-1.6$.  We choose to
fit a function that asymptotically approaches a constant at low \feh\
and increases linearly at large \feh.  A function that satisfies these
requirements is $\ln \left(e^{0.1} + e^{a + bx}\right)$.  The
least-squares fit with uniform weighting is

\begin{equation}
\label{eq:fehsyserr}
  \delta_{\mathrm{sys}}\mathfeh_{\mathrm{MRS}} =
  \ln \left(e^{0.1} + e^{-0.758 + 1.86\,\mathfeh}\right)\:.
\end{equation}

\noindent
The total error is

\begin{equation}
\label{eq:fehtoterr}
  \delta_{\mathrm{tot}}\mathfeh_{\mathrm{MRS}} =
  \sqrt{\left(\delta_{\mathrm{fit}}\mathfeh_{\mathrm{MRS}}\right)^2 +
    \left(\delta_{\mathrm{sys}}\mathfeh_{\mathrm{MRS}}\right)^2}\:.
\end{equation}

\noindent
The error bars on $\mathfeh_{\mathrm{MRS}}$ for all plots and tables
in this article are calculated from Eqs.~\ref{eq:fehsyserr} and
\ref{eq:fehtoterr}.

\subsection{Total Error on \afe}
\label{sec:afetoterr}
We repeat the same procedure to find the systematic error in
$\mathafe_{\mathrm{MRS}}$.  However, we compute the deviation from the
$\mathafe_{\mathrm{HRS}}$---accounting for HRS error---rather than the
mean of $\mathafe_{\mathrm{MRS}}$.

\begin{equation}
\mathrm{rms}\,\left(\frac{\mathafe_{\mathrm{MRS}} -
  \mathafe_{\mathrm{HRS}}}{\sqrt{\left(\delta_{\mathrm{fit}}\mathafe_{\mathrm{MRS}}\right)^2 + \left(\delta_{\mathrm{sys}}\mathafe_{\mathrm{MRS}}\right)^2 +
  \left(\delta\mathafe_{\mathrm{HRS}}\right)^2}}\right) = 1
\end{equation}

\noindent
The value that satisfies this equation is
$\delta_{\mathrm{sys}}\mathafe_{\mathrm{MRS}} = 0.03$.  The total
error is

\begin{equation}
\label{eq:afetoterr}
  \delta_{\mathrm{tot}}\mathafe_{\mathrm{MRS}} =
  \sqrt{\left(\delta_{\mathrm{fit}}\mathafe_{\mathrm{MRS}}\right)^2 +
    \left(\delta_{\mathrm{sys}}\mathafe_{\mathrm{MRS}}\right)^2}\:.
\end{equation}

\noindent
The error bars on $\mathafe_{\mathrm{MRS}}$ for all plots and tables
in this article are calculated from Eq.~\ref{eq:afetoterr}.

\subsection{Errors from Atmospheric Parameters}
\label{sec:atmerr}
Errors in \teff\ and \logg\ have the potential to change the measured
abundances significantly.  The total $\mathfeh_{\mathrm{MRS}}$ error
estimates in \S\,\ref{sec:fehtoterr} account for random error about
the true values of \teff\ and \logg, but not systematic offsets.  The
total $\mathafe_{\mathrm{MRS}}$ error estimates in
\S\,\ref{sec:afetoterr} do account for both random and systematic
error in the atmospheric parameters to the extend that \teff\ and
\logg\ in the comparison HRS studies are accurate.

\begin{deluxetable}{lcc}
\tablecolumns{3}
\tablewidth{0pt}
\tablecaption{Errors from Atmospheric Parameters\label{tab:atmerr}}
\tablehead{\colhead{Atmospheric Error}  &  \colhead{$\delta\mathfeh$}  &  \colhead{$\delta\mathafe$}}
\startdata
$\mathteff-250~\mathrm{K}$ & $-0.21 \pm 0.08$ & $+0.01 \pm 0.12$ \\
$\mathteff-125~\mathrm{K}$ & $-0.10 \pm 0.04$ & $+0.01 \pm 0.08$ \\
$\mathteff+125~\mathrm{K}$ & $+0.11 \pm 0.05$ & $-0.03 \pm 0.06$ \\
$\mathteff+250~\mathrm{K}$ & $+0.20 \pm 0.08$ & $-0.07 \pm 0.10$ \\
$\mathlogg-0.6$            & $-0.06 \pm 0.06$ & $+0.00 \pm 0.07$ \\
$\mathlogg-0.3$            & $-0.03 \pm 0.04$ & $+0.01 \pm 0.06$ \\
$\mathlogg+0.3$            & $+0.04 \pm 0.03$ & $-0.01 \pm 0.05$ \\
$\mathlogg+0.6$            & $+0.07 \pm 0.07$ & $-0.01 \pm 0.07$ \\
\enddata
\end{deluxetable}

We quantify the abundance error introduced by errors in atmospheric
parameters by varying \teff\ and \logg\ for all stars in the GC
sample.  We recompute abundances at $\mathteff \pm 125~\mathrm{K}$,
$\mathteff \pm 250~\mathrm{K}$, $\mathlogg \pm 0.3$, and $\mathlogg
\pm 0.6$.  Table~\ref{tab:atmerr} shows the differences in \feh\ and
\afe\ between the altered and unaltered atmospheres.  The numbers
presented are the mean difference and standard deviations for all
stars in all seven GCs.

As expected, increasing \teff\ causes \feh\ to increase because the
synthetic atmosphere must have a higher density of absorbers to
compensate for the temperature-induced weakening in line strength.
Increasing \logg\ also causes \feh\ to increase because a higher
electron pressure increases the density of $\mathrm{H}^-$ ions.
$\mathrm{H}^-$ is the dominant source of continuous optical and
near-infrared opacity in these cool giants.  Therefore, as \logg\
increases, the decreasing ratio of line opacity to continuous opacity
depresses line strength.  The synthetic atmosphere needs to be more
metal-rich to compensate.  We also point out a few trends with \teff,
\logg, and \feh:

\begin{enumerate}
 \item $\delta\mathfeh_{\mathrm{MRS}}$ is gradually less sensitive to
 $\delta\mathteff$ as \teff\ increases.
 \item $\delta\mathfeh_{\mathrm{MRS}}$ does not show a trend with
 $\delta\mathlogg$ as \logg\ increases.
 \item $\delta\mathfeh_{\mathrm{MRS}}$ does not show a trend with
 $\delta\mathteff$ as \feh\ increases.
 \item $\delta\mathfeh_{\mathrm{MRS}}$ is gradually more sensitive to
 $\delta\mathlogg$ as \feh\ increases.
 \item On average, $\delta\mathafe_{\mathrm{MRS}}$ has the same sign
 as $\delta\mathteff$ for $\mathteff < 4600~\mathrm{K}$ and the
 opposite sign otherwise.  $\delta\mathafe_{\mathrm{MRS}}$ is most
 sensitive to $\delta\mathteff$ at low \teff.
 \item $\delta\mathafe_{\mathrm{MRS}}$ does not show a trend with
 $\delta\mathlogg$ as \logg\ increases.
 \item $\delta\mathafe_{\mathrm{MRS}}$ does not show any trends with
 \feh.
\end{enumerate}

\subsection{Effect of Noise}
\label{sec:noise}
Stars at the tip of the RGB in MW GCs are easy targets for high
resolution spectrometers.  The intended targets of the MRS method are
much fainter.  Therefore, we explore the effect of noise on the
measurement of \feh\ and \afe.  To estimate S/N, we compute the
absolute deviation from 1.0 of all pixels in the continuum regions
(see \S\,\ref{sec:continuum}).  We clip pixels that exceed 3 times
this mean deviation.  The S/N per pixel is the inverse of the mean
absolute deviation of the remaining pixels.  To convert to S/N per
\AA, multiply by 1.74, the inverse square root of the pixel scale.

\begin{figure}
\plotone{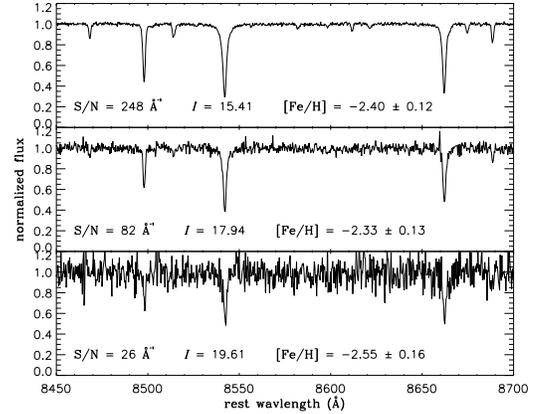}
\caption{Examples of the \ion{Ca}{2} triplet region of spectra at
  three different signal-to-noise ratios.  These are three different
  stars in NGC~2419, obtained with 20~minutes of exposure time on
  \deimos.  Shown in each panel is the S/N, $I$ magnitude, and
  measured \feh.  The value from one star (not shown) observed with
  HRS is $\mathfeh_{\mathrm{HRS}} = -2.32$
  \citep{she01}.\label{fig:examples}}
\end{figure}

Figure~\ref{fig:examples} shows the \ion{Ca}{2} triplet region of
three \deimos\ spectra of stars in \ngcb\ in three different S/N
regimes.  The three strongest lines are the \ion{Ca}{2} triplet, and
the five weaker lines are individual or blended \ion{Fe}{1}
transitions.  We show the \ion{Ca}{2} triplet to demonstrate spectral
quality, but we do not use any spectral information from any of the
three lines because we do not model them accurately.  Instead, we use
the weaker metal lines, any one of which is difficult to identify in
low S/N spectra.  As an ensemble over $\sim 2800$~\AA\ of spectral
range, they provide accurate abundances.  Below, we discuss spectra at
$\mathrm{S/N} \sim 10~\mathrm{\AA}^{-1}$, in which even the
\ion{Ca}{2} triplet is barely identifiable.

\begin{figure}
\plotone{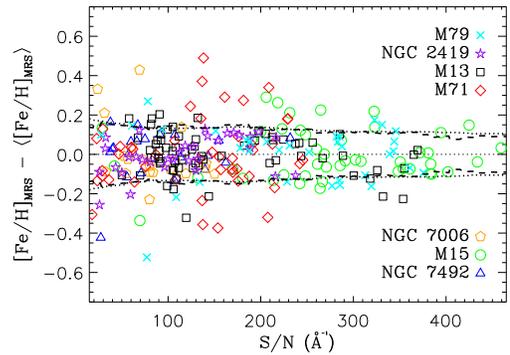}
\caption{The difference between individual measurements of
  $\mathfeh_{\mathrm{MRS}}$ and the cluster's mean
  $\langle\mathfeh_{\mathrm{MRS}}\rangle$ vs.\ signal-to-noise ratio.
  The dashed line is the rms in a moving window with a half width of
  $50~\mathaa^{-1}$.  The dotted line is the average error determined
  from Eqs.~\ref{eq:fehsyserr} and \ref{eq:fehtoterr} in the same
  moving window.\label{fig:sn1}}
\end{figure}

Figure~\ref{fig:sn1} shows the difference between
$\mathfeh_{\mathrm{MRS}}$ for an individual star and the mean
$\mathfeh_{\mathrm{MRS}}$ that we measure for its cluster
($\Delta\mathfeh_{\mathrm{MRS}}$) vs.\ S/N.  We also include the rms
of all points in a moving window with a half width of
$50~\mathaa^{-1}$.  At $\mathrm{S/N} > 200~\mathaa^{-1}$, the rms
scatter in $\Delta\mathfeh$ is about 0.15~dex.  The rms rises to about
0.20~dex in the range $20~\mathaa^{-1} < \mathrm{S/N} <
200~\mathaa^{-1}$.  Errors such as these are about the magnitude of
typical HRS abundance errors.  We also include the average
$\delta_{\mathrm{tot}}\mathfeh_{\mathrm{MRS}}$ in the same moving
window.  The close agreement between the two error averages
demonstrates that we determine
$\delta_{\mathrm{tot}}\mathfeh_{\mathrm{MRS}}$ well.

\begin{figure}
\plotone{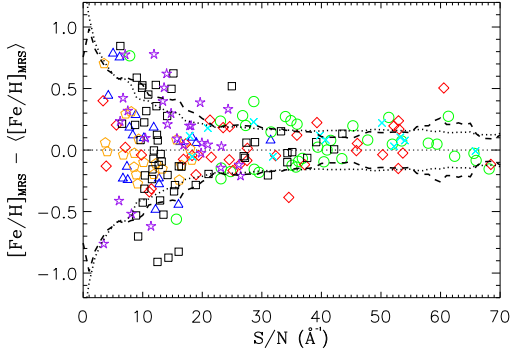}
\caption{Same as Fig.~\ref{fig:sn1} for measurements injected with
  artificial noise.  The ranges of both axes are different from
  Fig.~\ref{fig:sn1}.  The dashed line is the rms in a moving window
  with a half width of $5~\mathaa^{-1}$.  The dotted line is the
  average error determined from Eqs.~\ref{eq:fehsyserr} and
  \ref{eq:fehtoterr} in the same moving window.\label{fig:sn2}}
\end{figure}

In order to better scrutinize noisier spectra at $\mathrm{S/N} <
20~\mathaa^{-1}$, we inject the raw spectra with Gaussian random noise
proportional to the square root of the measured variance in each
pixel.  On average, the S/N of a given spectrum decreases by a factor
of 10.  We repeat the complete analysis, including continuum
determination and velocity cross-correlation.  Figure~\ref{fig:sn2}
shows that rms scatter gradually increases as S/N decreases, but even
at $\mathrm{S/N} \sim 10~\mathaa^{-1}$, the rms scatter is only
0.5~dex.  The \ion{Ca}{2} triplet is barely identifiable, and the
analysis includes only lines weaker than the triplet, yet the spectra
still contain enough information to yield decent metallicity
estimates.

\begin{figure}
\plotone{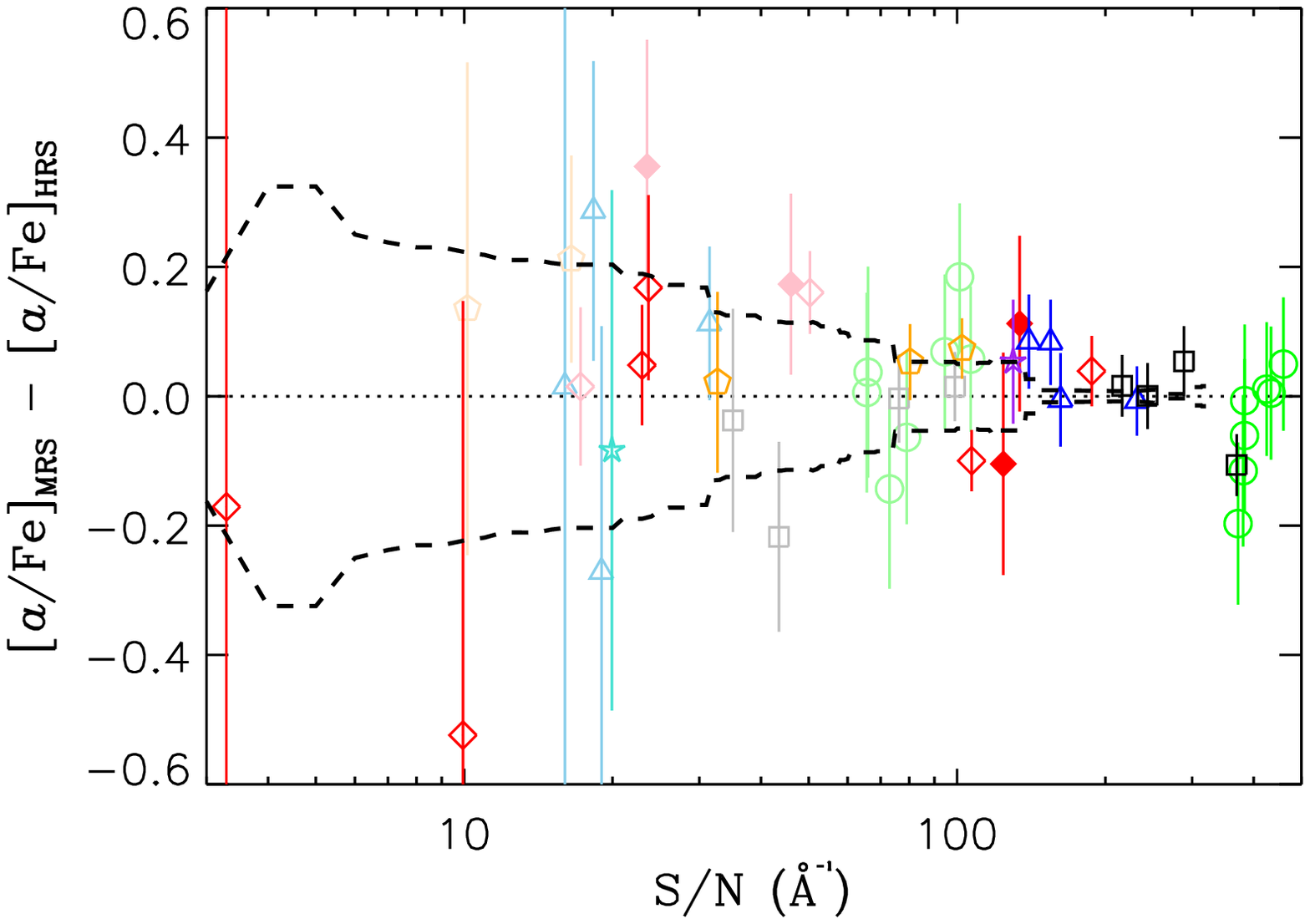}
\caption{Differences between $\mathafe_{\mathrm{MRS}}$ and
  $\mathafe_{\mathrm{HRS}}$ vs.\ signal-to-noise ratio.  The error
  bars are the sum in quadrature of
  $\delta_{\mathrm{tot}}\mathafe_{\mathrm{MRS}}$
  (Eq.~\ref{eq:afetoterr}) and $\delta\mathafe_{\mathrm{HRS}}$
  (Eq.~\ref{eq:afeerr_hrs}).  The dashed line is the average of
  $\sqrt{(\mathafe_{\mathrm{MRS}} - \mathafe_{\mathrm{HRS}})^2 -
  (\delta\mathafe_{\mathrm{HRS}})^2}$ in a moving window with a half
  width of 0.5~dex in S/N.  At $\mathrm{S/N} \ga 320$,
  $\delta\mathafe_{\mathrm{HRS}}$ completely accounts for the total
  error, and the argument of the square root becomes negative.  The
  symbols, colors, and references are the same as in
  Fig.~\ref{fig:teff} except that faded symbols represent measurements
  from spectra injected with artificial noise.\label{fig:snalpha}}
\end{figure}

We also investigate the effect of noise in estimating \afe.  We repeat
the comparison in Fig.~\ref{fig:alpha_avg} in which the sample is
restricted to those stars also observed with HRS.  In
Fig.~\ref{fig:snalpha} we plot the difference between
$\mathafe_{\mathrm{MRS}}$ and $\mathafe_{\mathrm{HRS}}$ vs.\ S/N.  In
the same figure, we also plot the results from the same stars injected
with artificial noise (faded symbols).  The dashed line is the scatter
about zero in a moving window with a half width of 0.5~dex in S/N,
correcting for the error on $\mathafe_{\mathrm{HRS}}$:
$\sqrt{(\mathafe_{\mathrm{MRS}} - \mathafe_{\mathrm{HRS}})^2 -
(\delta\mathafe_{\mathrm{HRS}})^2}$.  At $\mathrm{S/N} \ga 320$,
$\delta\mathafe_{\mathrm{HRS}}$ becomes comparable to
$\left|\mathafe_{\mathrm{MRS}} - \mathafe_{\mathrm{HRS}}\right|$.
Even at $\mathrm{S/N} \sim 10~\mathaa^{-1}$ (the typical S/N of a
one-hour exposure of an RGB star in \andromeda), the average error is
$\sim 0.2$~dex, small enough to distinguish between the halo plateau
value of $\mathafe \sim +0.3$ and $\mathafe \sim 0.0$.

A Gaussian noise model is a simplified representation of the actual
noise model.  In reality, the spectral error contains a systematic
component, particularly around night sky lines.  A thorough noise
model is beyond the scope of this article, but we intend to address this
issue in future work.  In the meantime, the Gaussian noise model gives
a good approximation of the abundance precision we expect with lower
S/N spectra.


\section{Applications}
\label{sec:applications}
GCs are already well studied at high resolution.  MRS abundance
measurement is not intended to provide more accurate abundances for
GCs.  After all, it has only two dimensions of abundance, although
future versions may even allow estimates of individual element
enhancements, such as Al, O, Na, Mg, Ca, Ti, La, Ba, and Eu.  However,
the real strength of MRS is to probe large samples at large distances.

Upcoming studies will focus on the metallicities and $\alpha$
enhancements of the dwarf galaxies of the MW and \andromeda.  We also
intend to explore the differences between the chemical properties of
different kinematic components of \andromeda: the cold dSphs, the
disrupted satellites and streams, and the hot halo.  We intend for the
MRS measurements to be a direct test of chemical evolution
simulations, such as those of \citet{fon06}.

In exploring these diverse systems, we will need to recognize some
shortcomings of the MRS technique.  First, we do not model TiO
absorption, and we have discarded all stars that show TiO.  In future
applications, we intend to continue discarding stars with $\mathteff <
4000$~K due to the complexity of modeling TiO absorption.  Second, the
errors at the higher metallicity of \ngcd\ \citep[$\mathfeh =
-0.76$,][]{pri05} become somewhat large (0.2~dex), though not larger
than typical errors for photometric or \ion{Ca}{2} triplet-based
metallicities.  Both problems will affect metal-rich samples such as
the inner halo of \andromeda\ \citep[$\langle\mathfeh\rangle =
-0.47$,][]{kal06}.  Finally, samples of very faint stars may require
spectral coaddition in bins of \teff\ and \logg\ to achieve S/N ratios
high enough to make measurements of $\alpha$ enhancement with
precision greater than 0.3~dex.  Although enhancements of individual
stars would be unrecoverable, coadded spectra would contain enough
information to estimate a stellar population's overall chemical
properties.

We have demonstrated that metallicity measurements of medium
resolution spectra via spectral modeling approach the accuracy and
precision of measurements from HRS.  Furthermore, the leverage of a
large number of absorption lines from many different elements allows
abundance measurement in multiple dimensions.  Presently, the method
estimates $\alpha$ enhancement, but future versions may provide
individual element enhancements.  Spectral modeling is superior to
spectrophotometric indices, such as the \ion{Ca}{2} triplet EW.
First, the precision from modeling is comparable or better at the same
S/N.  Second, modeling requires no empirical calibration.  Therefore,
the result is not tied to intrinsic parameters (e.g., [Ca/Fe]) of the
calibrators.  Additionally, modeling is not subject to the range of
metallicities of the calibrators.  Whereas empirical calibrations to
GCs will be inaccurate at $\mathfeh \la -2.4$, the MRS method is
applicable to arbitrarily low metallicities, which may exist in some
of the lowest mass MW satellites \citep{sim07}.  These advantages make
medium resolution spectral modeling the only abundance technique that
will be able to test theories of hierarchical structure formation
through large samples of individual stars at large distances.

\acknowledgments

The authors gratefully acknowledge P.~B.~Stetson for providing
photometry of all of the spectroscopic targets in this article and
J.~Simon and M.~Geha for providing the \deimos\ observations of
\ngca\ and \ngcb.  We thank R.~Kraft, D.~Lai, C.~Rockosi, I.~Ivans,
and M.~Shetrone for extremely useful discussions.

We acknowledge National Science Foundation grants AST-0607708,
AST-0307966, and AST-0607852 and NASA/STScI grants GO-10265.02 and
GO-10134.02.  ENK is supported by a NSF Graduate Research Fellowship.
Data herein were obtained at the W.~M. Keck Observatory, which is
operated as a scientific partnership among the California Institute of
Technology, the University of California, and NASA.  The Observatory
was made possible by the generous financial support of the W.~M. Keck
Foundation.  The analysis pipeline used to reduce the \deimos\ data
was developed at UC Berkeley with support from NSF grant AST-0071048.

{\it Facility:} \facility{Keck:II (DEIMOS)}

\clearpage

\begin{landscape}
\begin{deluxetable}{cccccccccccl}[p]
\tablecolumns{12}
\tablewidth{0pt}
\tabletypesize{\scriptsize}
\tablecaption{Abundances for Individual Stars\label{tab:mrsdata}}
\tablehead{\colhead{$\alpha$ (2000.0)}  &  \colhead{$\delta$ (2000.0)}  &  \colhead{$B$\tablenotemark{a}}  &  \colhead{$V$\tablenotemark{a}}  &  \colhead{$I$\tablenotemark{a}}  &  \colhead{\teff}  &  \colhead{\logg}  &  \colhead{$\mathfeh_{\mathrm{MRS}}$}  &  \colhead{$\mathafe_{\mathrm{MRS}}$}  &  \colhead{$\mathfeh_{\mathrm{HRS}}$}  &  \colhead{$\mathafe_{\mathrm{HRS}}$\tablenotemark{b}}  &  \colhead{HRS reference}}
\startdata
\cutinhead{\ngcc}
$16^{\mathrm{h}} 40^{\mathrm{m}} 52.1^{\mathrm{s}}$ & $+36 \arcdeg 29 \arcmin 25 \arcsec$ & $17.31 \pm 0.04$ & $16.64 \pm 0.03$ & $15.82 \pm 0.01$ & 5336 & 3.09 & $-1.69 \pm 0.13$ & $+0.33 \pm 0.07$ &      \nodata     &      \nodata     & \nodata        \\
$16^{\mathrm{h}} 40^{\mathrm{m}} 58.4^{\mathrm{s}}$ & $+36 \arcdeg 26 \arcmin 04 \arcsec$ & $16.70 \pm 0.02$ & $16.03 \pm 0.01$ & $15.17 \pm 0.01$ & 5204 & 2.80 & $-1.86 \pm 0.11$ & $+0.41 \pm 0.05$ &      \nodata     &      \nodata     & \nodata        \\
$16^{\mathrm{h}} 41^{\mathrm{m}} 01.6^{\mathrm{s}}$ & $+36 \arcdeg 28 \arcmin 00 \arcsec$ & $17.37 \pm 0.04$ & $16.79 \pm 0.02$ & $15.95 \pm 0.01$ & 5260 & 3.13 & $-1.71 \pm 0.12$ & $+0.28 \pm 0.07$ &      \nodata     &      \nodata     & \nodata        \\
$16^{\mathrm{h}} 41^{\mathrm{m}} 02.6^{\mathrm{s}}$ & $+36 \arcdeg 26 \arcmin 16 \arcsec$ & $15.68 \pm 0.01$ & $14.92 \pm 0.01$ &      \nodata     & 4965 & 2.26 & $-1.65 \pm 0.12$ & $+0.22 \pm 0.04$ & $-1.44 \pm 0.05$ & $+0.22 \pm 0.03$ & \citet{coh05a} \\
$16^{\mathrm{h}} 41^{\mathrm{m}} 04.8^{\mathrm{s}}$ & $+36 \arcdeg 27 \arcmin 45 \arcsec$ & $17.48 \pm 0.04$ & $16.92 \pm 0.01$ & $16.10 \pm 0.01$ & 5318 & 3.20 & $-1.82 \pm 0.12$ & $+0.23 \pm 0.09$ &      \nodata     &      \nodata     & \nodata        \\
$16^{\mathrm{h}} 41^{\mathrm{m}} 05.8^{\mathrm{s}}$ & $+36 \arcdeg 26 \arcmin 30 \arcsec$ & $17.47 \pm 0.04$ & $16.86 \pm 0.01$ & $16.05 \pm 0.01$ & 5379 & 3.19 & $-1.68 \pm 0.13$ & $+0.27 \pm 0.07$ &      \nodata     &      \nodata     & \nodata        \\
$16^{\mathrm{h}} 41^{\mathrm{m}} 06.2^{\mathrm{s}}$ & $+36 \arcdeg 25 \arcmin 22 \arcsec$ & $15.30 \pm 0.01$ & $14.46 \pm 0.01$ &      \nodata     & 4829 & 2.02 & $-1.65 \pm 0.12$ & $+0.23 \pm 0.04$ & $-1.44 \pm 0.05$ & $+0.17 \pm 0.04$ & \citet{coh05a} \\
$16^{\mathrm{h}} 41^{\mathrm{m}} 06.5^{\mathrm{s}}$ & $+36 \arcdeg 28 \arcmin 14 \arcsec$ & $14.29 \pm 0.01$ & $13.25 \pm 0.01$ &      \nodata     & 4522 & 1.37 & $-1.65 \pm 0.12$ & $+0.11 \pm 0.04$ & $-1.63 \pm 0.06$ &      \nodata     & \citet{sne04}  \\
$16^{\mathrm{h}} 41^{\mathrm{m}} 09.7^{\mathrm{s}}$ & $+36 \arcdeg 26 \arcmin 45 \arcsec$ & $15.52 \pm 0.02$ & $14.73 \pm 0.01$ & $13.79 \pm 0.01$ & 4997 & 2.19 & $-1.67 \pm 0.12$ & $+0.30 \pm 0.04$ &      \nodata     &      \nodata     & \nodata        \\
$16^{\mathrm{h}} 41^{\mathrm{m}} 09.9^{\mathrm{s}}$ & $+36 \arcdeg 27 \arcmin 42 \arcsec$ & $17.31 \pm 0.03$ & $16.73 \pm 0.02$ & $15.91 \pm 0.01$ & 5352 & 3.13 & $-1.52 \pm 0.14$ & $+0.13 \pm 0.08$ &      \nodata     &      \nodata     & \nodata        \\
\enddata
\tablecomments{Table~\ref{tab:mrsdata} is published in its entirety in
  the electronic edition of the Astrophysical Journal.  A portion is
  shown here for guidance regarding form and content.}
\tablenotetext{a}{P.~B.~Stetson has generously provided this
  photometry, which is preliminary pending his own publication of
  these data.}
\tablenotetext{b}{See \S\,\ref{sec:afe} for a discussion on
  $\mathafe_{\mathrm{HRS}}$ measurements of \citet{sne97},
  \citet{sne00}, and \citet{sne04}.}
\end{deluxetable}
\clearpage
\end{landscape}

\end{document}